\documentclass[aps,prd,endfloats*]{revtex4}
\usepackage{graphicx,amssymb,amsfonts}
\newcommand{\be}{\begin{equation}}
\newcommand{\ee}{\end{equation}}
\newcommand{\ba}{\begin{eqnarray}}
\newcommand{\ea}{\end{eqnarray}}

\begin{document}
\title{Muon Simulations for Super-Kamiokande, KamLAND and CHOOZ}
\author{Alfred Tang and Glenn Horton-Smith}
\affiliation{
Physics Department, Kansas State University, Manhattan, KS 66506}
\author{Vitaly A. Kudryavtsev}
\affiliation{
Department of Physics and Astronomy, University of Sheffield, Sheffield
S3 7RH, United Kingdom}
\author{Alessandra Tonazzo}
\affiliation{
APC and Universit\'e Paris 7, 75005 Paris, France}
\date{\today}

\begin{abstract}
Muon backgrounds at Super-Kamiokande, KamLAND and CHOOZ are calculated using
\texttt{MUSIC}.  A modified version of the Gaisser sea level muon distribution
and a well-tested Monte Carlo integration method are introduced.  Average muon
energy, flux and rate are tabulated.  Plots of average energy and angular
distributions are given.  Implications for muon tracker design in future
experiments are discussed.
\end{abstract}
\maketitle

\section{Introduction}
Muons are one of the most significant sources of background for underground
experiments.  An accurate and efficient numerical method to calculate the muon
rate and average energy at an underground lab is indispensable for this type of
research.  This work was originally motivated by a need to resolve the
question of the average muon energy for Daya Bay and KamLAND.  Since
Super-Kamiokande (Super-K) is essentially next to KamLAND and already has
many publications quoting its muon rates, it easily becomes an ideal source
of cross checks.  At the same time, the need to understand the
cosmic background at the far site of Double Chooz also arises.  Muon data from
the first CHOOZ experiment are subsequently made
available so that comparison with simulation becomes possible.  Given the
diversity of the experimental sites being discussed, some
effort is made to present the analysis in a more general context.  It is
hoped that the method presented here will be useful to a larger community.

The muon rate can be measured in an experiment by a number of methods. 
Measurement of muon energy on the other hand is quite
difficult.  Since a measurement made in one site under
a certain hill profile is unlikely to be transferable to another site,
an economical calculational method is the only practical solution.
For these reasons, whenever the average muon energy is needed for the
calculation of cosmogenic background rate, accurate
Monte Carlo simulation is often the most reasonable alternative.
Traditionally there have been some discrepancies in various reports regarding
muon rate and average energy for both Super-K and KamLAND.  For example,
different values of muon rates have been reported by different collaborators of
Super-K such as 1.88~Hz~\cite{koshio}, 2.2~Hz~\cite{fukuda},
2.5~Hz~\cite{gando} and 3~Hz~\cite{habig}.  
Some of these discrepancies are due to the differences of detector regions
or different selection criteria used in making various cuts.
For example, a cut at 1.6~GeV is made to eliminate the muon background in the
study of the upward throughgoing muons in Reference~\cite{fukuda}.  This cut
has the effect of lowering the cosmic muon rate.  In
Reference~\cite{gando}, the 2.5~Hz cosmic muon rate quoted is an
estimate used to make the spallation cut.  Differences in cosmic muon rates
due to the differences in detector regions
will be analyzed by simulation studies later.  As far as KamLAND is
concerned, accurate simulations of the average muon energy, flux and rate are
presently needed to aid the data analysis process.
In addition, the design of muon tracker systems for future experiments depend
on detailed simulations that can handle complicated topography.  In an effort
to build a reliable tool for all these needs, this paper introduces a complete
numerical method from the ground up---beginning with an improved
Gaisser sea level muon parameterization, showing in detail the logic of
the numerical method, making mention of useful numerical tools and ending with
numerous cross checks with experimental data including those of ground level
muons.

\section{A Bottom-Up Design}\label{soft}
\subsection{Preliminaries}
The goal of this section is to lay the theoretical foundation for how to
incorporate \texttt{MUSIC} with a user-supplied sampling algorithm.
Details of the implementation of the numerical method outlined in this
section can be found in Appendix~\ref{td}.
\texttt{MUSIC} is a \texttt{FORTRAN} subroutine that simulates the 
3-dimensional transport of muons through a slant depth $X$ of a material
taking into account energy loss due to ionization, pair production,
Bremsstrahlung and inelastic scattering~\cite{ku0,ku1}.  \texttt{MUSIC} is
composed of two main parts---(1) the Monte Carlo simulation of muon energy loss
and (2) the Monte Carlo simulation of angular deviation and lateral
displacement.  In
order to distinguish quantities related to initial muons on the surface and
the final muons that survive at a certain depth underground, the
subscripts $\mu0$ and $\mu$ will be used to denote the two types of muons
respectively.

The testing of the present numerical method involves the comparison of the
simulated results against published experimental and
simulated data.  The most convenient item of comparison is the vertical
muon intensity $I^v_\mu(h)$ versus vertical depth $h$ underneath a flat
surface in standard rock because of the abundance of experimental data.
In order to set the stage for the following discussions, several conventional
quantities are defined in the beginning.  For instance, $\theta$ is defined to
be the zenith angle of the line of sight of the muons and $\phi$ is the
azimuthal angle of the same, measured from the easterly direction
in the counter-clockwise sense.
Directional muon intensity $I_\mu(h,\cos\theta)$ has units of
$\rm cm^{-2}\,sr^{-1}\,s^{-1}$.  Vertical muon intensity is taken
as
\be
I^v_\mu(h)=I_\mu(h,\,\cos\theta=1),
\ee
which also has the unit of $\rm cm^{-2}\,sr^{-1}\,s^{-1}$.  Integrated muon
intensity is defined as~\cite{rossi}
\be
J_\mu(h)=\int_\Omega I_\mu(X,\,\cos\theta)\,d\Omega,
\label{ii}
\ee
where $\Omega$ defines the solid angle coverage and the slant depth $X$ is
the distance
traveled by the muon in rock.  The argument $X$ in $I_\mu(X,\,\cos\theta)$
is generally a function of $\cos\theta$.  For example, $X=h/\cos\theta$ in the
case of a flat surface.  The unit of integrated muon intensity
is $\rm cm^{-2}\,s^{-1}$.

\subsection{Modified Gaisser Parameterization}
The accuracy of the simulation depends on \texttt{MUSIC},
the parameterization of the surface muon intensity and the user-supplied
sampling algorithm.  A standard atmospheric muon parameterization is 
given by Gaisser~\cite{gaisser} as
\be
{dN_{\mu0}\over dE_{\mu0} d\Omega}
\simeq A{0.14 E^\gamma_{\mu0}\over{\rm cm^2\,\,sr\,\,s\,\,GeV}}
\left\{{1\over1+{1.1\tilde E_{\mu0}\cos\theta\over115}}+
{.054\over1+{1.1\tilde E_{\mu0}\cos\theta\over850}}+r_c\right\}.
\label{gai}
\ee
Muon energy $E_{\mu0}$ at the surface is measured in GeV and
$\theta$ is the angle subtended between the incoming cosmic ray
particle and the normal to the upper atmospheric layer.  If the earth were
flat, $\theta$ is also the zenith angle on the ground surface.  Since the earth
is not flat, a correlation needs to be made between the zenith angle on the
ground surface and the angle measured on the upper atmosphere.  In order to
clarify the distinction between the two definitions of angle, a new symbol
$\theta^\star$ is invented to denote the angle measured on the upper atmosphere
as a function of $\theta$ which is assigned specifically to the zenith angle on
the ground surface from now on.  The calculation
of $\theta^\star$ will be explained later.  The symbols $A$, $\gamma$ and $r_c$
refer to the overall scale factor, power index and the ratio of the prompt
muons to pions respectively.  In the low energy regime, $E_{\mu0}$ needs to
be modified slightly by an energy loss through the atmosphere.  The
symbol $\tilde E_{\mu0}$ denotes the energy of muon on top of the atmosphere.
The differentials of time $dt$ and area $dA$ are omitted from the denominator
on the left-hand-side of Eq.~(\ref{gai}) for the sake of simplicity.  The
original Gaisser parameterization has $A=1$, $\gamma=2.70$,
$\tilde E_{\mu0}=E_{\mu0}$ and $r_c=0$.  For large depth greater than
1--2~km~w.e. (kilometer water equivalence), the LVD parameterization~\cite{lvd}
is recommended.  In that case, $A=1.84$, $\gamma=2.77$.
Since this work primarily concerns simulations for relatively shallow depths
as in Super-K, KamLAND and CHOOZ, the Gaisser parameterization is adequate
for the high energy part ($E_{\mu0}>(100/\cos\theta)$~GeV) of the
spectrum.  Since there are enough low energy muons that survive at shallow
depths, rare high energy muons ($E_{\mu0}>10^6$~GeV) are omitted from the
calculations.  The valid energy range for the Gaisser parameterization is
$(100/\cos\theta)<E_{\mu0}<10^6$~GeV and small angle $\theta<70^\circ$ where
the effect due to the curvature of the earth is negligible.  In the low energy
limit ($E_{\mu0}\le(100/\cos\theta)$~GeV), the Gaisser parameterization is
significantly higher than the observed values.  The expected angular
dependence of $\cos^n\theta$ with $n\sim2$ in this regime must also be taken
into account.  To satisfy all these additional requirements in the small
$E_{\mu0}$ and large $\theta$ regimes, the following modifications to
Eq.~(\ref{gai}) are suggested for
$(1/\cos\theta^\star)<E_{\mu0}<(100/\cos\theta^\star)$~GeV:
\ba
r_c&=&10^{-4},\label{rc}\\
\Delta&=&2.06\times10^{-3}\left({950\over\cos\theta^\star}-90\right),
\label{delta}\\
\tilde E_{\mu0}&=&E_{\mu0}+\Delta,\label{te}\\
A&=&1.1\left({90\sqrt{\cos\theta+0.001}\over1030}\right)^
{4.5\over E_{\mu0}\cos\theta^\star}.
\label{a}
\ea
It is important to note the term involving $\cos\theta$ inside the square
root of Eq.~(\ref{a}) does not have a star.
The LVD publications set the upper limit on the ratio of prompt muons to
pions to be $r_c<2\times10^{-3}$ at 95\% confidence level~\cite{lvd,lvd2}.
However $\chi^2$ of the fits is lower for smaller values of $r_c$ such that
the choice of Eq.~(\ref{rc}) is justified by statistical reason.
The symbol $\Delta$ in Eq.~(\ref{delta}) has the interpretation of
mean energy loss of muons in the atmosphere.  The value $2.06\times10^{-3}$
refers to the stopping power of matter against muons in units of GeV per
$\rm g/cm^2$ at $E_{\mu0}\simeq50$~GeV where the radiative effects reach
1\%~\cite{bichsel}.  The multiplication of this value with the mean muon
slant depth in the atmosphere will give the mean energy loss of muons in the
atmosphere.  A commonly quoted value of the atmospheric height $h_F$ is
1000~$\rm g/cm^2$.  Reference~\cite{dar} quotes a specific value of
$h_F=1030$~$\rm g/cm^2$ along with a value of interaction length (the
average distance between the point where a primary proton enters the atmosphere
and the point where a muon is produced) $\lambda_N=120$~$\rm g/cm^2$.  The
atmospheric height $h_F$ is a function of scale height $h_0$ which in
turn is a function of temperature.  In addition the stopping power used in
Eq.~(\ref{delta}) is simply a rough estimate.  For these reasons, $h_F$
should be adjusted to produce the best fit of experimental data in the low
energy regime.  For the purpose of constructing $\Delta$, the choice of
$h_F=950$~$\rm g/cm^2$ is made.  Beside the aforementioned value of
interaction length $\lambda_N=120$~$\rm g/cm^2$ quoted in Reference~\cite{dar},
many other values have also been quoted in the literatures, {\it e.g.}
$\lambda_N=77.6$~$\rm g/cm^2$~\cite{elbert} and
$\lambda_N=80$~$\rm g/cm^2$~\cite{honda}.  Again for the purpose
of constructing the fits, a median value of $\lambda_N=90$~$\rm g/cm^2$ is
chosen for $\Delta$.  Putting all these values together, the mean muon slant
depth is $(950/\cos\theta^\star-90)$~$\rm g/cm^2$ such that the final form of
Eq.~(\ref{delta}) is obtained.  For low energy muons, there is a slight
difference, $\Delta$, between the muon energy at ground level $E_{\mu0}$ and
the muon energy on top of the atmosphere $\tilde E_{\mu0}$.
Since the critical energy (the threshold by which the
mechanism changes from radiation losses to ionization losses)
used in Eq.~(\ref{gai}) refers to $\tilde E_{\mu0}$,
an adjustment is needed for the low energy regime as given in Eq.~(\ref{te}).
The meaning of Eq.~(\ref{a}) is essentially the multiplication of an
effective factor of 1.1 due to the nuclear enhancement of
multiplicity~\cite{dar} and the probability of muon decay $S_\mu$.  The
form of $S_\mu$ used in Eq.~(\ref{a}) is similar to that in
Reference~\cite{dar} and can easily be derived as follows: the decay
probability is related to decay length $L=-v\gamma\tau\ln R$ such that
$S_\mu=\exp(-\lambda_F/L)$ where $\lambda_F\equiv h_F/\cos\theta^\star$ is
the slant height of the atmosphere, $v$ is muon velocity,
$\gamma\equiv1/\sqrt{1-v^2}$ is the Lorentz factor, $\tau$ is the muon lifetime
and $0\le R\le1$ is a uniformly distributed number~\cite{ingelman}.  In
the muon energy regime $E_{\mu0}>(1/\cos\theta^\star)$~GeV,
the approximation $v\simeq1$ is acceptable.  Furthermore, if the choice
of $R=\lambda_N/\lambda_F$ is made along with the standard substitutions
$\gamma=E_{\mu0}/m_\mu$ and $h_F m_\mu/\tau\simeq1.04$~GeV, the original
form of $S_\mu$ in Reference~\cite{dar} is recovered.  This derivation reveals
that the form of $S_\mu$ in Reference~\cite{dar} does not incorporate any
matter and geomagnetic effects which may be important for low energy muons.
In the present work, nonlinear effects are taken into consideration by
assuming a modification to the decay length to achieve the best fits of
experimental data such that
\be
\tilde L\equiv 0.231\ln\left({\sqrt{\cos\theta+0.001}\over\cos\theta^\star}
\right)L.
\label{tl}
\ee
At this point, Eq.~(\ref{tl}) is purely phenomenological.  There is no simple
physical explanation for this change other than the fact that it fits the
low energy muon data.
By replacing $L$ with $\tilde L$ in Eq.~(\ref{tl}) and repeating the derivation
of $S_\mu$ above, Eq.~(\ref{a}) is obtained.

The modifications in Eqs.~(\ref{rc})--(\ref{a}) alone cannot fit the data in
the lowest energy range.  For $E_{\mu0}\le1/\cos\theta^\star$~GeV, the basic
form of the parameterization is the same as Eqs.~(\ref{gai})--(\ref{a}) with
the exception that the substitution
\be
E_{\mu0}\to{3E_{\mu0}+7\sec\theta^\star\over10}
\label{low}
\ee
is made before $E_{\mu0}$ is passed to Eqs.~(\ref{gai})--(\ref{a}).
The substitution in Eq.~(\ref{low}) is
just another phenomenological tool to achieve good fits with experimental data.

\begin{figure}
\centering
\includegraphics[scale=0.55]{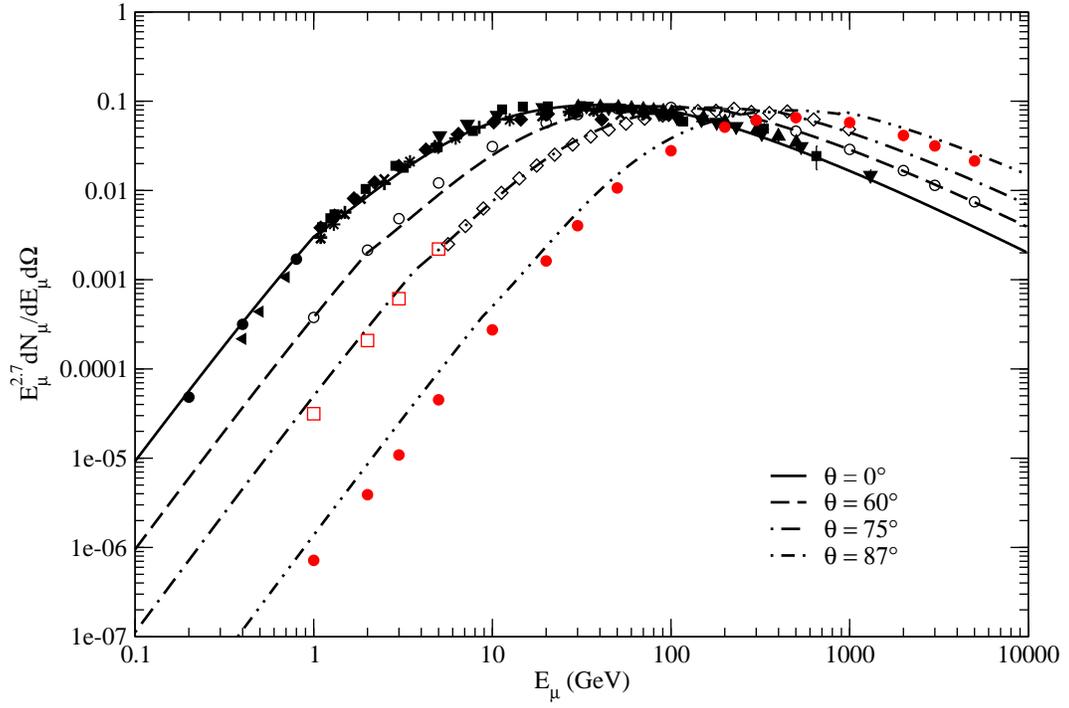}
\caption{\label{gaif}
Fits of the modified Gaisser parameterization to experimental data in the low
energy regime between $\theta=0$ and $\theta=87^\circ$.  The experimental data
are taken from References~\cite{gaisser}, \cite{pascale}--\cite{hayman}.
The modified Gaisser parameterization is given by
Eqs.~(\ref{gai})--(\ref{a}) and (\ref{low})--(\ref{tstar}).}
\end{figure}

The value of $\cos\theta^\star$ is sometimes calculated using a simple
geometrical extrapolation assuming that the altitude of first interaction is
known {\it a priori}.  The present work takes a different
approach by using a more complicated extrapolation method described in
Reference~\cite{chirkin} that shows how $\cos\theta^\star$
can be extracted from an integral equation by equating interaction length
$X(\theta)=X(0)$.  In essence the formula below taken from
Reference~\cite{chirkin} parameterizes the numerical solution of the integral
equation:
\be
\cos\theta^\star=\sqrt{x^2+p_1^2+p_2x^{p_3}+p_4x^{p_5}\over
1+p_1^2+p_2+p_4},
\label{tstar}
\ee
where $x\equiv\cos\theta$, $p_1=0.102573$, $p_2=-0.068287$, $p_3=0.958633$,
$p_4=0.0407253$ and $p_5=0.817285$.  The terms involving $\cos\theta$ in
Eq.~(\ref{gai}) must be replaced by $\cos\theta^\star$ for consistency.
Eq.~(\ref{a}) is protected against division by zero because
$\cos\theta^\star\ge0.103458$ for $\cos\theta\ge0$ according to
Eq.~(\ref{tstar}).  The modified Gaisser
parameterization is based on the world data set and hence represent
an average of the global sea level muon distribution.  The geomagnetic field
affects only the low energy spectrum, typically below 2~GeV for integrated muon
intensity~\cite{bhat} and approximately less than 20~GeV for vertical muon
intensity~\cite{motoki}.  The east-west effect is also shown to be negligible
at ground level~\cite{hansen} by careful simulations.  For the purpose of
calculating the muon overburden deep underground, geomagnetic effects can be
ignored because low energy sea level muons will not survive through rock by
default.  In essence, the present parameterization is composed of the
union of 3 segments:  $E_{\mu0}>(100/\cos\theta^\star)$~GeV
(the standard Gaisser formula),
$(1/\cos\theta^\star)<E_{\mu0}\le(100/\cos\theta^\star)$~GeV
(Eqs.~(\ref{rc})--(\ref{a}))
and $E_{\mu0}\le(1/\cos\theta^\star)$~GeV (Eq.~(\ref{low})).
Figure~\ref{gaif} illustrates the quality of the fits between the modified
Gaisser formula and experimental data.  The goodness of fit tends to degrade
only at very large angles $(\theta>85^\circ)$.  The worst disagreement between
experimental data and the parameterization in those cases is about 40\%.
However the worst case scenario of the 40\% disagreement
occurs only at low energy ($E_{\mu0}<10$~GeV) and a relatively small sector
at large angles ($\theta>85^\circ$)
so that the integrated spectrum is dominated by the very accurate parts of the
parameterization at smaller angles.   Finally it should be emphasized
that the modifications to the low energy part of the standard Gaisser
formula outlined in Eqs.~(\ref{rc})--(\ref{a}) and (\ref{low}) will not have
any significant
impact on the simulations of deep underground experiments.  Nevertheless an
accurate description of the low energy part of the sea level muon
distribution is important for calculating the muon overburden for sites
situated at shallow depths such as the near sites of the Double Chooz
and Daya Bay experiments.

\subsection{Modeling Physical Observables}
Directional intensity at depth $h$ is obtained by integrating the
Gaisser parametrization and the muon survival probability over the initial
muon energy $E_{\mu0}$ at the surface as
\be
I_\mu(X,\cos\theta^\star,\phi)=\int^\infty_0 dE_{\mu0}\,
P(E_{\mu0},X,\theta^\star,\phi)\,
{dN_{\mu0}(E_{\mu0},\cos\theta^\star)\over dE_{\mu0}\, d\Omega}.
\label{i}
\ee
The probability function $P(E_{\mu0},X,\theta^\star,\phi)$ defines
the survival probability of a muon with initial energy $E_{\mu0}$ traversing
a slant depth $X$ from the zenith angle $\theta^\star$ and the azimuthal angle
$\phi$.  It is emphasized that the symbol $I_\mu$ as used in this paper has
a different meaning than $I_\mu$ in Reference~\cite{ku3} in that the
latter refers to a differential muon intensity containing a probability
distribution function $P(E_\mu,E_{\mu0},X,\theta^\star,\phi)$ which is related
to $P(E_{\mu0},X,\theta^\star,\phi)$ as per
\be
P(E_{\mu0},X,\theta^\star,\phi)\equiv\int dE_\mu\,
P(E_\mu,E_{\mu0},X,\theta^\star,\phi).
\ee
In Reference~\cite{ku3}, $2\times10^7$ muons with energies from 0.1 up to
1000~TeV were propagated through 15~km.w.e.of rock.  The values of
$P(E_\mu,E_{\mu0},X)$ were stored and then
integrated numerically using an equation similar to Eq.~(\ref{i}) to obtain
energy and angular distributions at any particular depth.
This work takes a different
approach by evaluating $P(E_{\mu0},X,\theta^\star,\phi)$ {\em in situ}
in the Monte Carlo integration.  This approach requires a smaller
number of simulated events (typically $5\times10^6$) and is more
versatile when applied to arbitrary hill profiles when
$P(E_{\mu0},X,\theta^\star,\phi)$ must be re-evaluated every time
the $(x,y)$-coordinates are changed.  In principle the transport of muons
from the surface to a point underground and vice versa are equivalent
as far as the calculation of energy loss is concerned.  The
most important requirement of the present method is the uniform generation of
$E_{\mu0}$, $\theta$ and $\phi$ as shown in Appendix~\ref{td}.  An arbitrary
energy dependent observable ${\mathcal O}_\mu(E_\mu)$
can be estimated as
\be
\langle {\mathcal O}_\mu(X,\cos\theta^\star,\phi)\rangle=
{1\over I_\mu(X,\,\cos\theta^\star,\phi)}
\int^\infty_0 dE_{\mu0}\,
{dN_{\mu0}(E_{\mu0},\cos\theta^\star)\over dE_{\mu0}\, d\Omega}
\int^\infty_0 dE_\mu\,{\mathcal O}_\mu(E_\mu)\,
P(E_\mu,E_{\mu0},X,\theta^\star,\phi)\,
.
\label{e}
\ee
The bracketed quantity on the left-hand-side of Eq.~(\ref{e}) represents the
average of ${\mathcal O}_\mu$.  The bracket will be dropped from now on
for the sake of simplicity unless ambiguities arise due to the choice of
symbols.
With Eqs.~(\ref{i}) and (\ref{e}), vertical intensity and average energy
are
\ba
I_\mu^v(h)&=&\int^\infty_0 dE_{\mu0}\,
P(E_{\mu0},h,\theta^\star=0,\phi)\,
{dN_{\mu0}(E_{\mu0},\cos\theta^\star=1)
\over dE_{\mu0}\, d\Omega},
\label{iv}\\
E_\mu^v(h)&=&{1\over I^v_\mu(h)}
\int^\infty_0 dE_{\mu0}\,
{dN_{\mu0}(E_{\mu0},\cos\theta^\star=1)\over dE_{\mu0}\, d\Omega}
\int^\infty_0 dE_\mu\,E_\mu\,P(E_\mu,E_{\mu0},h,\theta^\star=0,\phi).
\label{ev}
\ea
There are many different ways to implement Eqs.~(\ref{iv}) and (\ref{ev})
numerically.  Appendix~\ref{td} describes an efficient and accurate Monte Carlo
method.  Simulated values of $I^v_\mu(h)$ beneath a flat surface are compared
against experimental and simulated data in Figures~\ref{cr1}.  The results
obtained by using the modified Gaisser parameterization incorporating
Eqs.~(\ref{rc})--(\ref{a}) at low energy agrees with experimental data more
closely than those using the standard Gaisser parameterization only.
Figure~\ref{ratiov} shows the consistency between simulated and experimental
data at shallow depths.  (The interpretation of
Figure~\ref{ratiov} will be discussed more fully in Appendix~\ref{td}.)
The integrated muon intensity and average energy are
\ba
J_\mu&=&\int_\Omega d\Omega \,\int^\infty_0 dE_{\mu0}\,
P(E_{\mu0},X,\theta^\star,\phi)\,
{dN_{\mu0}(E_{\mu0},\cos\theta^\star)
\over dE_{\mu0}\, d\Omega},
\label{is}\\
\langle E_\mu\rangle&=&{1\over J_\mu}
\int_S d\Omega \,\int^\infty_0 dE_{\mu0}\,
{dN_{\mu0}(E_{\mu0},\cos\theta^\star)\over dE_{\mu0}\, d\Omega}
\int^\infty_0 dE_\mu\,E_\mu\,P(E_\mu,E_{\mu0},X,\theta^\star,\phi).
\label{es}
\ea
$J_\mu$ and $\langle E_\mu\rangle$
in Eqs.~(\ref{is}) and (\ref{es}) are functions of the
location of the point of sampling underneath a topographic profile.  The
arguments of these functions, namely the coordinates of the point of sampling,
are understood and therefore not displayed explicitly.  The brackets around
$E_\mu$ on the left-hand-side of Eq.~(\ref{es}) are dropped in the
following text whenever the reference to average muon energy is clear from
the context of the discussion.
Integrated intensity can be computed in a similar way as vertical intensity.
The only difference is that the depth $X$ is now a function of $\theta$ and
$\phi$.  In the case of a flat surface, the relation takes on a simple form
$X=h/\cos\theta$.  In the case of an arbitrary hill profile, there is no
longer any simple relationship among $X$, $\theta$ and $\phi$.

The key of the present numerical method is uniform generation of integration
variables which can be achieved
in reliable ways through various uniform generation algorithms such as
the \texttt{CERNLIB} routine \texttt{RANLUX}.  The logic of the method is
relatively simple so that there is little ambiguity of its correctness.  All
these observations lead to the conclusion that \texttt{MUSIC} is sufficiently
accurate over the relevant range of muon energy.  Simulated integrated muon
intensity and average energy are compared against published simulations in
Table~\ref{cr}.  

\section{Preparing the Calculation}
\subsection{Digital Maps}
\begin{figure}
\includegraphics[scale=0.6]{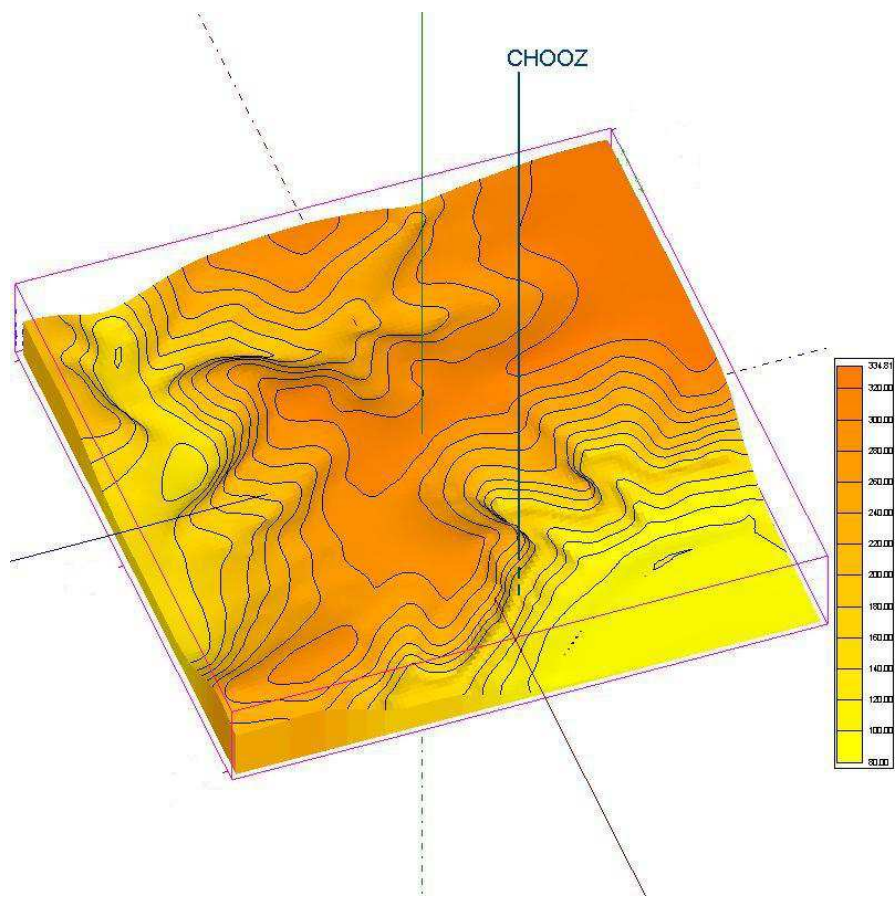}
\caption{\label{dcmap}
A visualization of the 3D topological profile of the Ardennes Mountains over
the CHOOZ site generated by \texttt{3DField} from a 2D contour map.}
\end{figure}

The starting point of a muon simulation over an arbitrary hill profile is the
digital map of the surrounding topology.  The accuracy of a digital map
directly affects the accuracy of the calculation so that a detailed knowledge
of the hill profile is important.  According to a contour map published in
Reference~\cite{suzuki}, KamLAND is separated from Super-K by 187m
and its bearing is N66.6E with respect of Super-K.  The top of the Super-K
tank and the bottom of the KamLAND tank are situated at 350m above sea level.
Both sites are almost directly underneath the peak of Ikenoyama
at 1.35~km.  Due to their proximity, both sites have very
similar muon energy and flux.  However the sizes of the two detectors are
vastly different so that their muon rates scale accordingly.  
The digital map of the Ikenoyama mountain profile around Super-K
is extracted from a code used by M. Nakahata originally to calculate the muon
background for Super-K.  The Super-K digital map sets its origin
at the location of the detector and parameterizes coordinates in terms of
$(\rho,\,\phi,\,h)$.  This particular format does not allow a simple
coordinate transformation of the origin from Super-K to KamLAND.
As a result, the digital map of KamLAND is constructed independently from a
contour map for this work.  In order to guarantee a sufficient solid
angle coverage for the simulation, both digital maps cover circular areas of
radius 3950~m.  The topological map of CHOOZ is generated from a 2D contour
map using a shareware called \texttt{3DField}~\cite{3dfield}.
A visualization of the digital
map over the Ardennes Mountains is shown in Figure~\ref{dcmap}.
\texttt{3DField} has the option of generating an \texttt{ASCII} data file
containing the $(x,\,y,\,z)$ coordinates of the latticized hill profile.
One side of the CHOOZ detector is beneath a steep hill so that a large range
of the solid angle coverage is parameterized by a relatively small set of
lattice points.  In order to increase the density of lattice points over
the steep section of the Ardennes hill profile, another digital map is
created over a smaller area around the detector.  At the end, both digital
maps are spliced together to form one single digital map so that the entire
solid angle coverage is represented more evenly.

\subsection{Detector Geometry}
The calculation of the average muon rate depends on the details of the
detector geometry.
For Super-K, the parameters that define the geometry of the cylindrical tank
are $L_0=41.40\,\rm m$ and $R_0=19.65\,\rm m$.  An inner volume is defined
to eliminate the simulations of very small muon track lengths
inside the detector geometry that do not intersect the active region of
the detector.  The choice of the inner volume
is not critical for the calculation of muon rate inside the outer tank
$R^t_\mu$.  For the purpose of this work, the inner volume used in the
simulation of $R^t_\mu$ is also the inner detector volume of Super-K whose
dimensions are
$L=36.20\,\rm m$ and $R=16.90\,\rm m$.  The inner and outer tanks of Super-K
have almost the same aspect ratios so that the muon rate inside the inner
tank $R^i_\mu$ can be obtained simply by scaling $R^t_\mu$ according to the
ratio of the physical areas $A_0$ of the two tanks.  The geometry of the
cylindrical
tank at KamLAND is defined by $L_0=19.68\,\rm m$ and $R_0=9.50\,\rm m$.  The
inner spherical volume of KamLAND for the purpose of this simulation is taken
to be the area bounded by the buffer region with $R=8.25\,\rm m$.  For
the simulation of the muon rate inside the KamLAND detector volume, a sphere
of radius $R_0=6.50\,\rm m$ is used.  In the case of CHOOZ, the cylindrical
tank has
the dimensions of $L_0=5.5\,\rm m$ and $R_0=2.75\,\rm m$.  The inner detector
is filled with Gd-loaded liquid scintillator and has the shape of a short
cylinder with hemispherical end caps.  Muon rate inside the Gd-loaded region
is not simulated in the present work.

\subsection{Rock Composition}
Chemical composition of the rock affects a \texttt{MUSIC} simulation in that
two out of three cross section files need to be calculated with specific
rock data {\it a priori}.  Table~\ref{chem} gives the chemical composition
of the Ikenoyama and Ardennes rock.  The average atomic number and weight are
$\langle Z\rangle=10.13$ and $\langle A \rangle=20.42$ for the Ikenoyama rock
and $\langle Z\rangle=11.8$ and $\langle A \rangle=24.1$ for the Ardennes rock.
Hydrogen composition is 2.2\% for Ikenoyama and negligible for Ardennes.
The rock density and the radiative length are $\rho=2.70\,\rm g/cm^3$ and
$\lambda=25.966\,\rm g/cm^2$ for Ikenoyama and $\rho=2.81\,\rm g/cm^3$ and
$\lambda=23.3\,\rm g/cm^2$ for Ardennes respectively.
The present simulation for CHOOZ takes the approximate chemical composition
and the average rock density~\cite{baldini} as inputs.
The Ardennes Mountains has a complicated rock density
profile with a layer of dense rock ($3.1$~$\rm g/cm^3$)~\cite{apollonio}
on the northeast sector.

In principle complex geological profiles can
be incorporated into the \texttt{MUSIC} simulation by a stratified approach in
which a simulation is segmented according to regions of different
densities, average atomic numbers $\langle Z\rangle$, average atomic masses
$\langle A\rangle$ and radiation lengths.  Although the stratified approach is
possible, it may not be easily achieved in practice.  Aside from the
computational challenge of simulating a complex geological profile, information
of the geological profile obtained by geological surveys may not be
generated with sufficient details to support a realistic simulation in the
first place.  Fortunately the stratified approach can be avoided in
many cases.  If $\langle Z\rangle$ and
$\langle A\rangle$ are constant and only density varies with depth, the mean
density should give the same average muon energy and flux as those generated
from stratified densities. Varying densities may affect the profiles of angular
distributions as in the CHOOZ case shown in Section~\ref{results}.  Radiation
length affects only the lateral displacement, which is not under investigation
in this paper. Small changes in $\langle Z\rangle$ and $\langle A\rangle$ (up
to 10\%) should not seriously affect the muon flux as long as the mean values
of all layers are found accurately.  This work does not attempt to simulate the
detailed geological profile of the Ardennes Mountains.  It is shown in
Section~\ref{results} that the simulated results due to the simplification of
the Ardennes geological profile are consistent with the previous CHOOZ
measurements within errors and that simulated results of a uniform
Ikenoyama mountain profile agree with experimental data.

\section{Results and Discussions}
\label{results}
\subsection{Average Muon Rate}
The calculation
of muon rate depends on the the effective area of the detector.  The basic
strategy of calculating the effective area $A$ is to multiply the physical
area $A_0$ with the ensemble average of the inner products of randomly
generated unit vectors $\hat r_i$ pointing from an inner volume and the unit
normal vectors $\hat r^i_0$ pointing away from the outer surface.  In this
case, the ensemble average also constrains the pseudo survival probability
of muons $\langle P\rangle$ that will be defined more precisely by
Eq.~(\ref{p}) in Appendix~\ref{td}.  Figure~\ref{vol} visualizes
how the inner products are done.  An intuitive way to think about the
effective area $\bar A$ is
\be
A={A_0\,\langle P\rangle\over N}\,\sum^N_{i=1} \hat r_i\cdot\hat r_{0i}.
\label{inner}
\ee
In the case of a cylindrical detector, the physical area is $A_0=\pi R^2_0+
2\pi R_0L_0$.  Similarly $A_0=4\pi R^2_0$ for a spherical detector and so on.
If Eq.~(\ref{inner}) is used, the average muon rate $R_\mu$ is simply
\be
R_\mu=J_\mu\,A.
\label{macro}
\ee
The average muon flux $J_\mu$ is always sampled at the center of the detector
volume in this work.  Although the macroscopic strategy defined by
Eqs.~(\ref{inner}) and (\ref{macro}) gives reasonable results, a microscopic
strategy to compute the muon rate is considered more
accurate, namely
\be
R_\mu=\int d{\mathbf A}\cdot\hat r\,
\int_\Omega d\Omega \,\int^\infty_0 dE_{\mu0}\,
P(E_{\mu0},X,\theta^\star,\phi)\,
{dN_{\mu0}(E_{\mu0},\cos\theta^\star)
\over dE_{\mu0}\, d\Omega},
\label{micro}
\ee
where $d\mathbf A$ is a differential area element along the the detector wall
and $\hat r$ is a unit vector along the muon line of sight.  Both are functions
of position along the detector wall, $\cos\theta^\star$ and $\phi$.
Appendix~\ref{td} gives a numerical implementation of Eq.~(\ref{micro}).

Table~\ref{tab}
summarizes the main results in terms of average muon energy, flux and
rate.  The muon rate in the outer tank of Super-K generated
by the present method is somewhere between the experimental values published
in References~\cite{gando} and \cite{fukuda}.
The 3~Hz muon rate quoted in Reference~\cite{habig} is most likely a rounded
figure.  It is not clear if the 1.88~Hz muon rate in Reference~\cite{koshio}
refers to the inner detector volume only.  If so,
it would agree with the simulated result very closely.
The muon rate at Kamiokande is usually quoted as 0.44~Hz~\cite{nakahata}.
Since KamLAND is slightly larger than Kamiokande, the muon rate should be
scaled according the ratio of the physical areas $A_0$ of the two tanks
which becomes approximately 0.5~Hz.  The unofficial measured rates on the
KamLAND outer detector and the balloon are 0.75 and .21 Hz respectively.
They
differ from the simulated results by about 10\% and 17\% respectively.
The official measured muon rate in the spherical buffer region
of radius $R_0=8.25$~m is 0.34~Hz~\cite{kl1} and the simulated result is
0.396~Hz (14\% difference).
The muon flux of 0.4~$\rm m^{-2}\,s^{-1}$ quoted by CHOOZ~\cite{apollonio} is
smaller than the simulated exact result by about 35\%, which is attributable
to the single digit of precision of the quoted rate and the approximated
geology used in our simulation.
The errors in the simulated results in Table~\ref{tab} are a combination of
the systematic error from map-making and the statistical error
from the Monte Carlo simulation.
The systematic error of the mountain profile coming from the calculation
of the scale that relates physical distance on the contour map to the relative
distance on the digital map is taken to be 0.5\%.
The systematic errors of
$E_\mu$, $J_\mu$ and $R_\mu$ are calculated by varying the slant depth $X$
by 0.5\% before passing it to \texttt{MUSIC}.
The statistical variance is calculated in the usual way by varying the
random seed.

\subsection{Energy and Angular Distributions}
\begin{figure}
\includegraphics[scale=0.55]{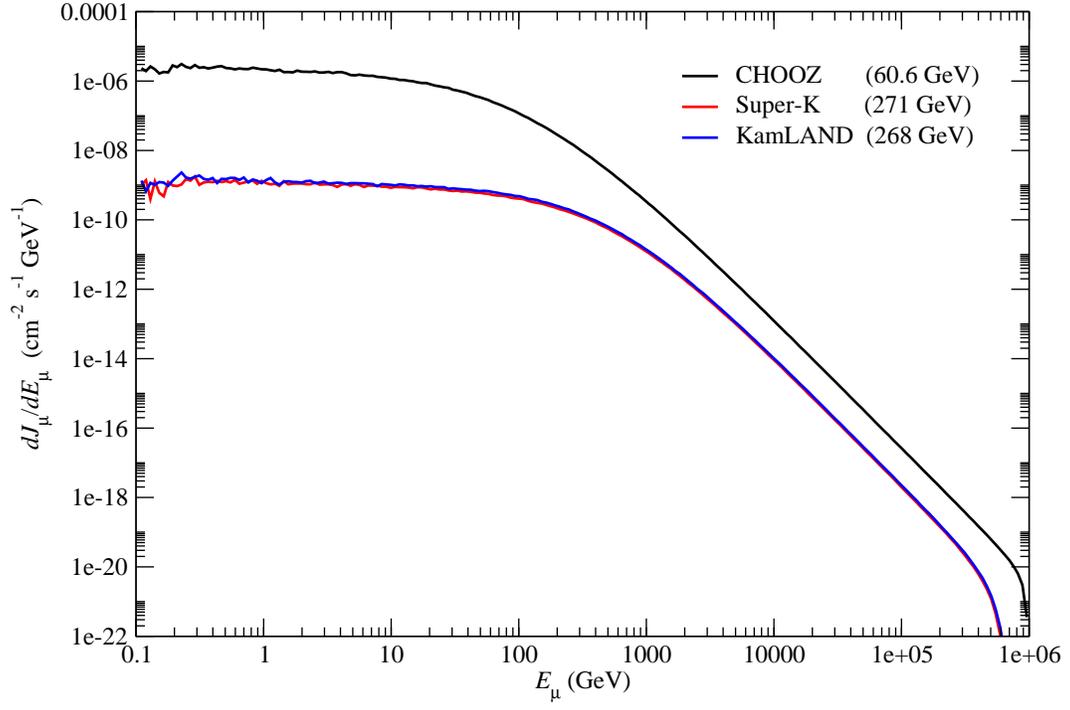}
\caption{\label{fkk}
Integrated muon intensity distribution at Super-K, KamLAND and CHOOZ.
The number of energy bins is $M=500$.  The total number of simulated events is
$5\times10^6$.  The average muon energies of the three sites are quoted
in the legend.}
\end{figure}

The energy distribution $dJ_\mu/dE_\mu$ in Figure~\ref{fkk} is defined by
the formula
\be
J_\mu\simeq
\int^\infty_0 dE_{\mu}\,{dJ_\mu\over dE_\mu},
\label{dist}
\ee
and has the unit of $\rm GeV^{-1}\,cm^{-2}\,s^{-1}$.  Angular and double
differential distributions can also be defined in similar ways.
Appendix~\ref{td} describes numerical implementations for various types of
distributions.  Figure~\ref{fkk} plots the cosmic muon energy distributions at
Super-K, KamLAND and CHOOZ.  The purpose of the figure is to show the global
properties of the energy distributions of various experiments.  Although the
distributions look smooth on the log-log scale, the fluctuations in the low
energy regime ($E_\mu<1$~GeV) will become more apparent on the semi-log scale.
Fortunately the fluctuation in the low energy part of the spectrum
on the log scale is suppressed by the smallness of the
Jacobian that contains a factor of $E_\mu$ so that the accuracy of
the calculations of the average muon energy $E_\mu$ and flux $J_\mu$ are
not affected.  If the energy distribution of stopping muons is needed,
generation of $E_{\mu0}$ and $\cos\theta$ according to the surface muon
distribution is recommended.

\begin{figure}
\includegraphics[scale=0.6,angle=270]{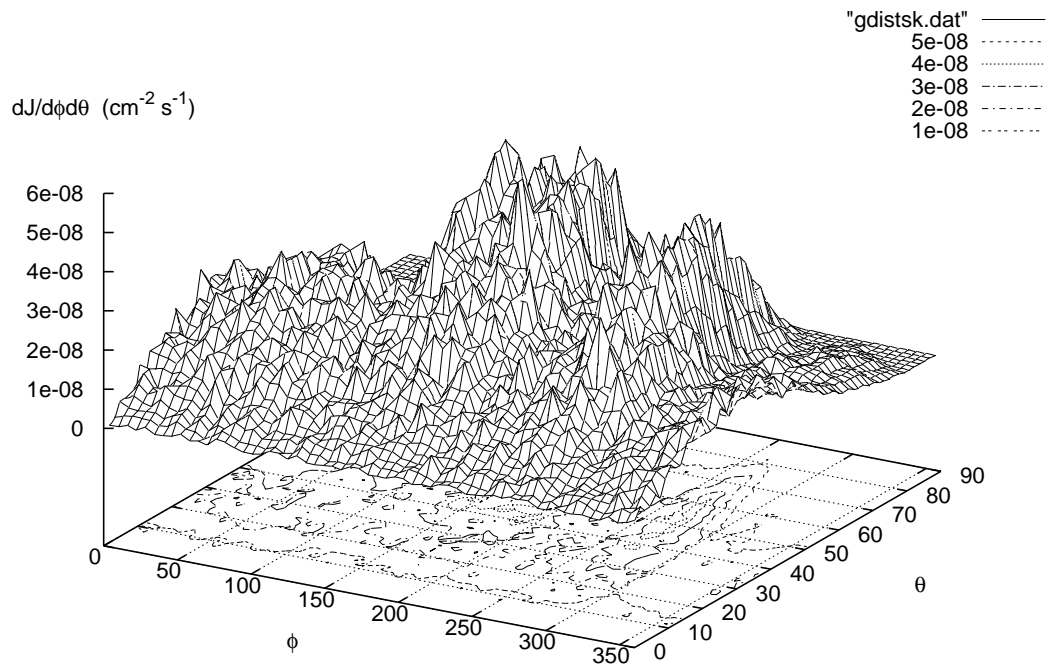}
\caption{\label{gsk}
Angular distribution of final muons at Super-K.  The total number of
simulated events is $5\times10^6$.  The Super-K digital map defines
$\phi=0$ to be along the northerly axis and the sense of rotation to be
clockwise.}
\end{figure}

\begin{figure}
\includegraphics[scale=0.6,angle=270]{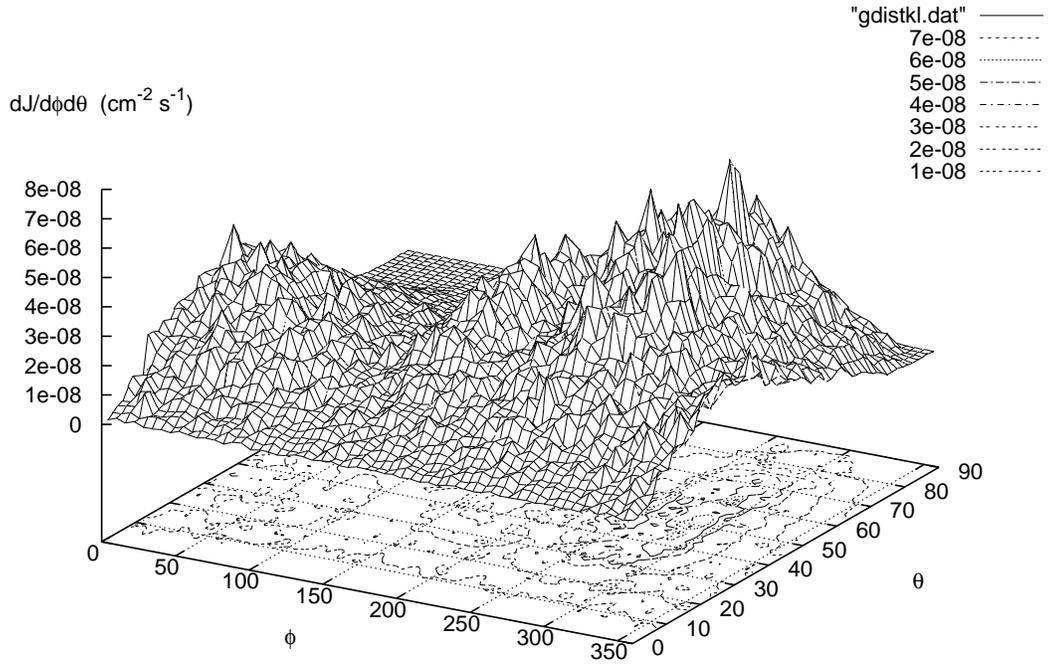}
\caption{\label{gkl}
Angular distribution of final muons at KamLAND.
The total number of simulated events is $5\times10^6$.}
\end{figure}

\begin{figure}
\includegraphics[scale=0.6,angle=270]{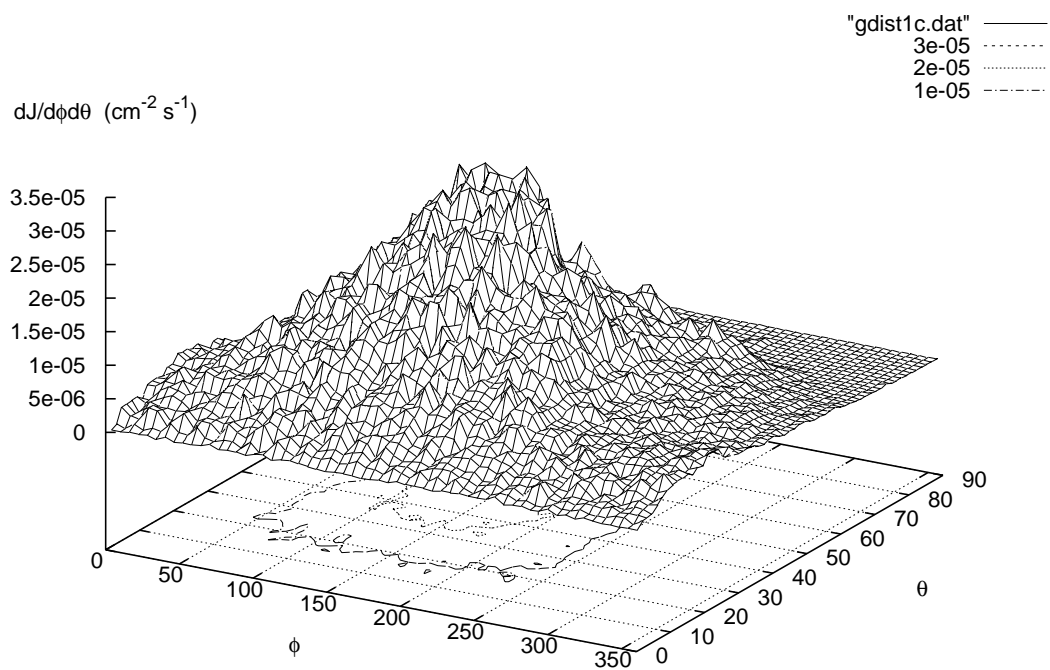}
\caption{\label{gdc}
Angular distribution of final muons at CHOOZ.
The total number of simulated events is $5\times10^6$.}
\end{figure}

\begin{figure}
\includegraphics[scale=0.55]{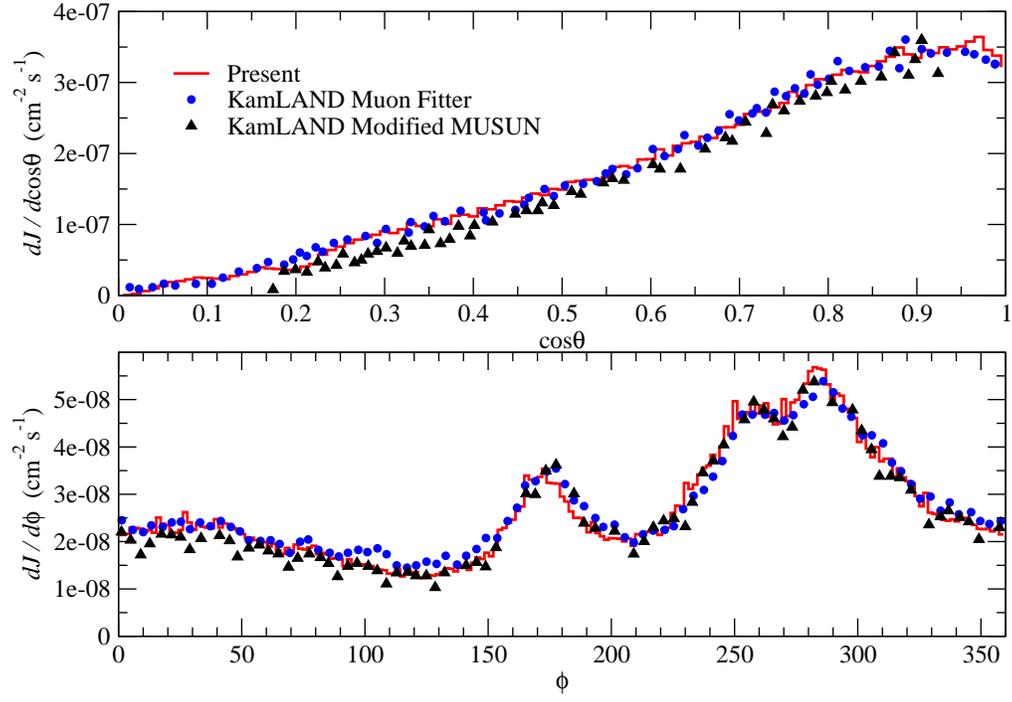}
\caption{\label{akk}
Comparisons of the $\cos\theta$ and $\phi$ distributions of final muons at
KamLAND.  The total number of simulated events of the present \texttt{MUSIC}
simulation is $5\times10^6$.  The muon fitter results~\cite{dwyer} represent
actual experimental data.  The modified \texttt{MUSUN} simulation~\cite{araki}
is a standard rock calculation while the present simulation is an exact
calculation
based on the Ikenoyama rock composition.  Both the muon fitter and modified
\texttt{MUSUN} results are rescaled from arbitrary units to fit the present
results.}
\end{figure}

\begin{figure}
\includegraphics[scale=0.55]{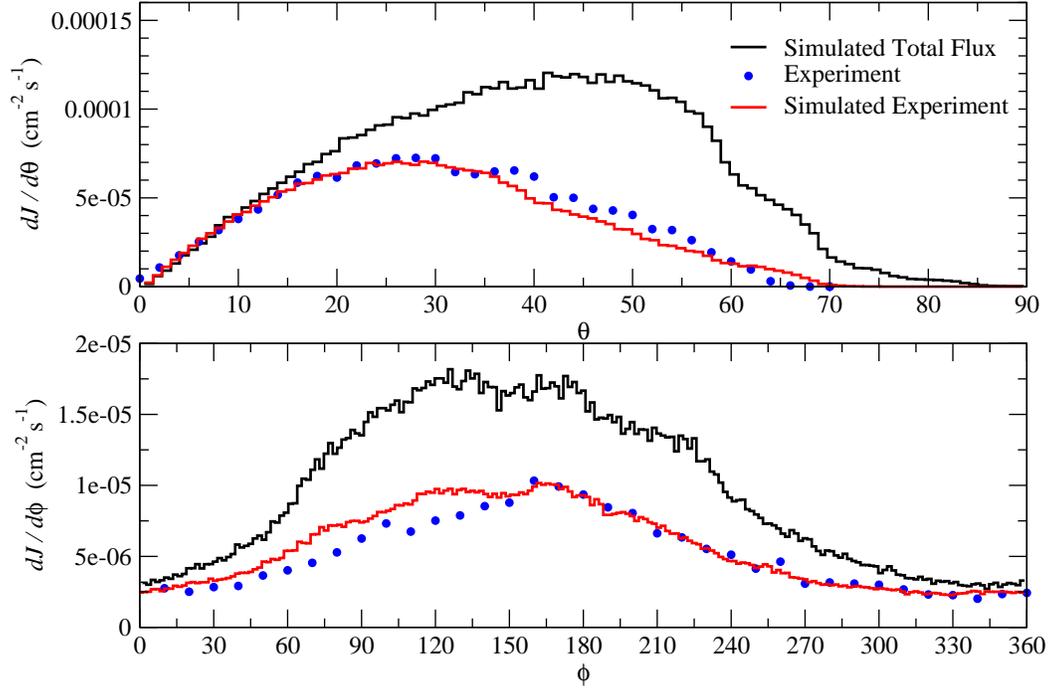}
\caption{\label{adc}
Comparisons of the $\theta$ and $\phi$ distributions of final muons at
CHOOZ.  The total number of simulated events of the present \texttt{MUSIC}
simulation is $5\times10^6$.  ``Simulated Total Flux'' refers to the simulated
muon flux integrated over the entire hemispherical solid angle.  The
total flux is binned to generate the angular distributions.
``Simulated Experiment'' is the simulated muon flux
integrated over a limited range of solid angle described by the geometry of
the RPC plates used in the cosmic ray experimental setup on the CHOOZ site.
Both the experimental and simulated experimental
results are rescaled such that the small angle parts of all three $\theta$
distributions agree.}
\end{figure}

Figures~\ref{gsk}--\ref{gdc} illustrate the
angular distributions of muons.
Experience shows that $5\times10^6$ simulated events are generally adequate to
generate reasonably good quality distributions in most cases.
The polar angle $\theta$ in the relevant plots is defined to be the
zenith angle consistent with Eqs.~(\ref{ii}) and (\ref{gai}).
The azimuthal angle $\phi$ is set to zero when the final muon
travels from east to west.  The momentum of the final muon is opposite to
the line of sight connecting the detector and the muon and is defined by
$\theta$ and $\phi$.  The only exception to the present definition of
$\theta$ and $\phi$ is
Figure~\ref{gsk} because the Super-K digital map uses a different convention.
Figure~\ref{akk} compares the $\cos\theta$ and $\phi$
distributions between simulations and experiment at KamLAND.  Figure~\ref{adc}
compares the $\theta$ and $\phi$ distributions between simulations and a cosmic
ray experiment done on the CHOOZ site in 1994.  The experiment consists of
four $1\times1$~$\rm m^2$ PRC (Resistive Plate Chambers)
plates separated from each other by 20~cm.  The
simulation of the experiment defines a muon event as the coincidence of any
two of the RPC plates being triggered.  The difference between the
simulated and the experimental $\theta$ distributions can be explained
by the fact that a significant number of the muons coming from the
steep section of the Ardennes hill profile cannot be detected by a muon
tracker composed of top and bottom horizontal plates only.  In order to
measure the muons coming from large zenith angles, additional trackers are
needed on the sides.  The remaining small differences between
the simulated and experimental results and the aberrations presumably arising
from the variation in geology described in Reference~\cite{apollonio} are
not simulated in this work.  The difference between the
simulated and experimental $dJ/d\phi$ for $0<\phi<150^\circ$ in
Figure~\ref{adc} is consistent with an unpublished result in an internal
note of the CHOOZ collaboration.  Notwithstanding the lack of detailed
treatment of smaller features, a macroscopic picture emerges by the way of a
qualitative comparison in the performance of two types of muon trackers
represented by the horizontal plate cosmic ray experiment on the CHOOZ site
and the muon veto system of KamLAND.
The simulated $\cos\theta$ and $\phi$
distributions agree well with KamLAND experimental data because the muon
tracker system of KamLAND
has full sensitivity over the entire range of the hemispherical solid angle.
The disagreement between the exact simulation of the $\theta$
distribution and the experimental data measured by horizontal plates
of the CHOOZ cosmic ray experiment in Fig.~\ref{adc} shows that the
contribution of muon flux from the sides cannot be neglected in the case of
a detector located underneath a hilly topology.  The obvious exception to this
claim will be the case where a detector is situated underneath a flat surface
so that the slant depth grows with $\sec\theta$.

\begin{figure}
\includegraphics[scale=0.55]{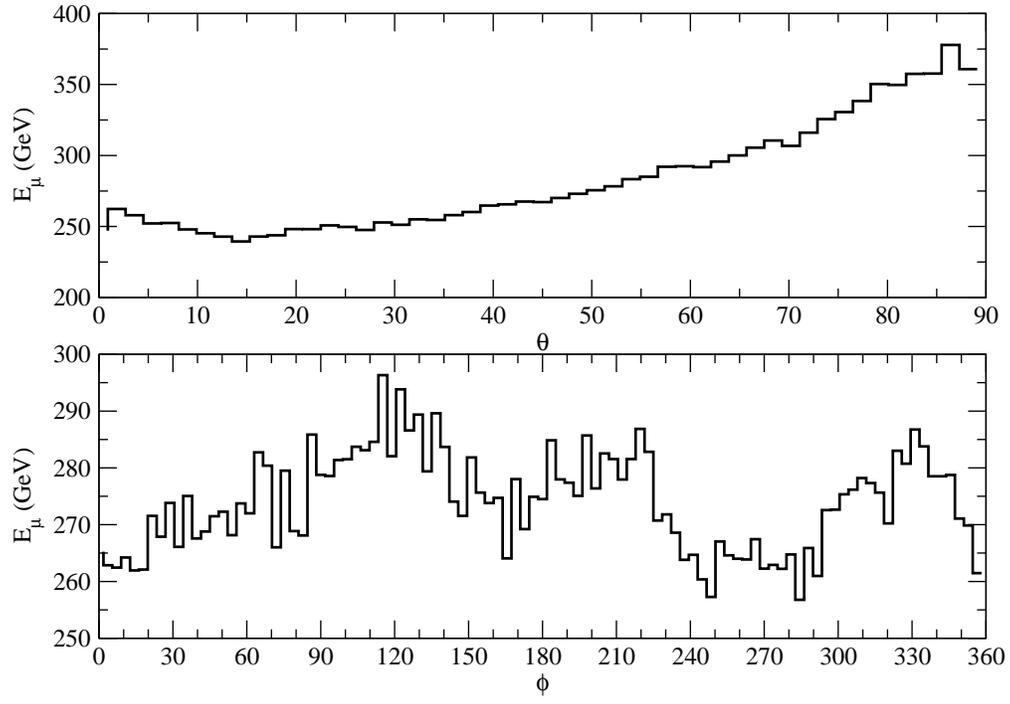}
\caption{\label{eptkl}
Average muon energy $E_\mu$ versus $\theta$ and $\phi$ at KamLAND.
The total number of simulated events is
$5\times10^6$.}
\end{figure}

\begin{figure}
\includegraphics[scale=0.55]{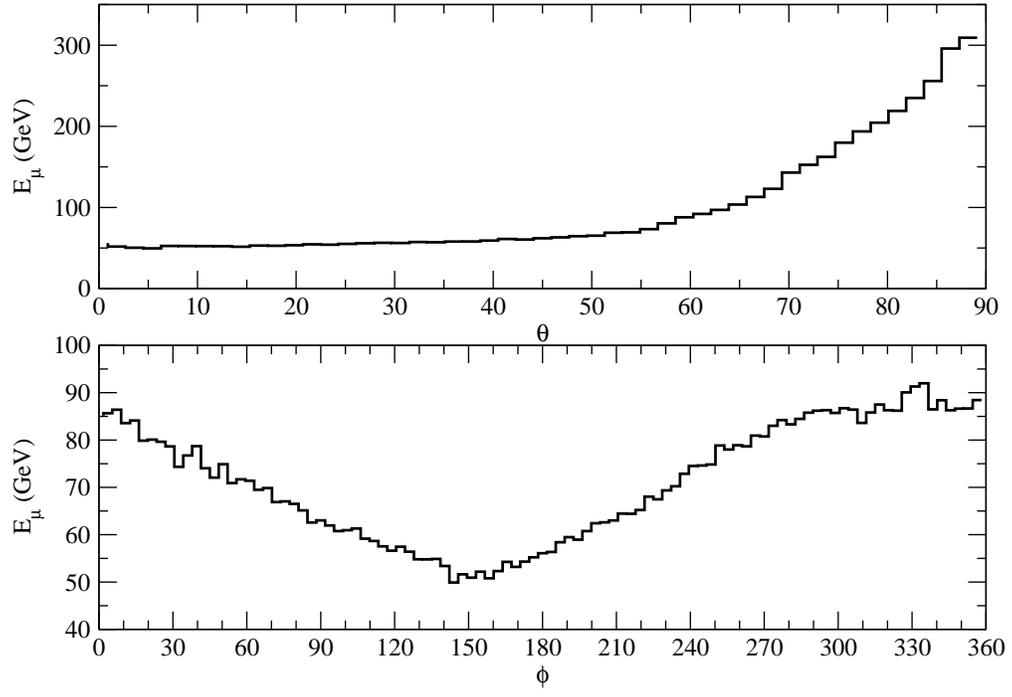}
\caption{\label{ept1c}
Average muon energy $E_\mu$ versus $\theta$ and $\phi$ at CHOOZ.
The total number of simulated events is
$5\times10^6$.}
\end{figure}

Figures~\ref{eptkl}--\ref{ept1c} plot the average muon energy versus $\theta$
and $\phi$ for KamLAND and CHOOZ respectively.  It is noted that the
differential flux in Figs.~\ref{akk} and \ref{adc} tends to vary inversely
with the
average muon energy $E_\mu$ per angle in Figs.~\ref{eptkl} and \ref{ept1c}.
This anti-correlation is intuitive in that average muon energy
generally increases with slant depth while muon intensity decreases with slant
depth.

\section{Conclusion}
The described method integrates \texttt{MUSIC}, a modified Gaisser
parameterization of the sea-level muon spectrum, and a uniform sampling
algorithm for the surface topography.  The method is efficient, robust,
and portable.  Given sufficiently accurate geological data, the method
is capable in principle of predicting muon rates and mean energies within a
few percent accuracy for depths less than 2.5~km.w.e., as indicated by the
error estimates in Table~\ref{tab}. 
In practice, simulations performed using simplified geology assuming
uniform rock composition lie within 10$\sim$20\% of observed rates published
by Super-Kamiokande and KamLAND, and within 35\% of the published flux at
the geologically more complex CHOOZ site.
Although muon simulations for any arbitrary hill profile have already been
done many times by other researchers previously,
there are very few complete documents
approximating a pedagogical introduction to the numerical method itself.
Although muon rates can
be measured in an experiment, muon energy is difficult to measure
so that knowledge of the average muon energy depends on simulation.  For this
reason, the reliability of the numerical method is very important.  In
applications such as the estimation of muon background in reactor
$\theta_{13}$ experiments, the method of uniform generation of variables
can serve as an additional cross check for accuracy. 
Although the standard Gaisser or LVD parameterizations are generally
adequate for the simulations of the deep underground experiments,
the modified Gaisser parameterization is indispensable for
shallow depth muon simulations.

\begin{acknowledgments}
The authors thank K. B. Luk, M. C. Chu, W. K. Ngai, M. Y. Guan, J. Cao,
Y. F. Wang, J. Learned, J. Shirai, K. Ichimura and J. Formaggio for providing
relevant information and helpful discussions.  Members of the
KamLAND and CHOOZ Collaborations have graciously released previously
unpublished data.  Special thanks go to D. Dwyer, S. Freedman, P. Decowski,
L. Hsu and G. Giannini and K. Inoue on this point.
A. Tang was supported by the
Chinese University of Hong Kong through the postdoctoral fellowship program
where part of this work was done.
V. A. Kudryavtsev wishes to thank Particle
Physics and Astronomy Research
Council (UK) and the EU FP6 Programme ILIAS (Contract
RII3-CT-2004-506222) for support.
G. Horton-Smith and A. Tang are supported by the Department of Energy
and the State of Kansas.
\end{acknowledgments}

\appendix
\section{Technical Details of the Numerical Method}\label{td}
The quality of the 3D topological map is crucial for an accurate calculation
of muon overburden.  It is usually the first and the most important step.
Care should be taken to
remove the disconnected parts of the mountain profile.  If a ray defined by a
specific set of $(\theta,\,\phi)$ passes through disconnected parts of
the mountain geometry resulting in several different values of slant depth
$X$, the smallest value of $X$ is used for that particular solid angle.

The range of the energy in the integral goes in principle from 0 to infinity.
It is more advantageous to numerically integrate over $\log E_{\mu0}$ instead
of $E_{\mu0}$.  (In this work, $\log$ refers to base 10 logarithm and $\ln$ to
base $e$.)  On the other hand, integration over $E_{\mu0}$ will give
essentially the same results.  The range of numerical integration over
muon energy $E_{\mu0}$ is labeled
by the lower and upper bounds, $E_l$ and $E_u$ respectively.  This work chooses
not to change the variable in such a way to integrate up to
$E_{\mu0}\to\infty$.  More specifically, the natural cut-off point ought to
be a sharp drop in the muon spectrum which in turn correlates with the knee
of the primary proton spectrum between 1000 and 10000~TeV.  The change in
the muon spectrum is 5--10 times lower than that so that a reasonable estimate
of the upper limit is $E_u=10^6$~GeV.  As a practical consideration, it is
more computationally efficient to set
$E_l$ not strictly as the rest mass of muon $m_\mu$ but the minimum muon energy
needed to survive the minimum slant depth of a particular geographic
profile so that CPU time is not wasted in simulating muons that cannot
survive by default.

After the
change of variables from $E$ to $\log(E)$, Eq.~(\ref{i}) is transformed as
\be
I_\mu(X,\cos\theta^\star,\phi)\simeq
\ln10\int^{\log E_u}_{\log E_l}
d\log E_{\mu0}\,P(E_{\mu0},X,\theta^\star,\phi)E_{\mu0}
{dN_{\mu0}(E_{\mu0},\cos\theta^\star)\over d E_{\mu0}\,
d\Omega}.
\label{il}
\ee
An integral in the Monte Carlo method~\cite{nr} can be approximated as
\be
\int^{y_2}_{y_1} f(x,y)dy_x\simeq
\langle f(x,y)\rangle\cdot(y_2-y_1),
\label{mc}
\ee
where $\langle f(x,y)\rangle$ is the average of $f(x,y)$ over $y$.
With Eqs.~(\ref{il}) and (\ref{mc}), Eq.~(\ref{i}) can be calculated
numerically as
\be
I_\mu(X,\cos\theta^\star,\phi)\simeq{\ln10\,(\log E_u-\log E_l)\over N}
\sum_{\{i\}}E_{\mu0i}\,
{dN_{\mu0}(E_{\mu0i},\cos\theta^\star_i)\over d E_{\mu0i}\, d\Omega}.
\label{im}
\ee
The symbol $\{i\}$ denotes a subset of simulated events corresponding to
surviving muons.  Information of $X$ and $\phi$ on the right-hand-side of
Eq.~(\ref{im}) are defined as inputs in the simulation and are not shown
formally.  The probability function
$P(E_{\mu0},X,\theta^\star,\phi)$ is not explicitly computed in
Eq.~(\ref{im}) by design.  The simplicity of this algorithm translates to
saving in memory.  Since
the probability function is not computed explicitly for each combination of
$E_{\mu0}$, $\theta$ and $\phi$, it is essential that the generation of these
integration variables is uniform so that the probability function is
calculated implicitly when the sum is divided by $N$ in Eq.~(\ref{im}).
As a test of the accuracy of present method, it will be shown later that the
uniform generation of integration variables gives exactly the same
results as those calculated by the Gaussian quadrature method in the case of
ground level muons.

\begin{figure}
\includegraphics[scale=0.55]{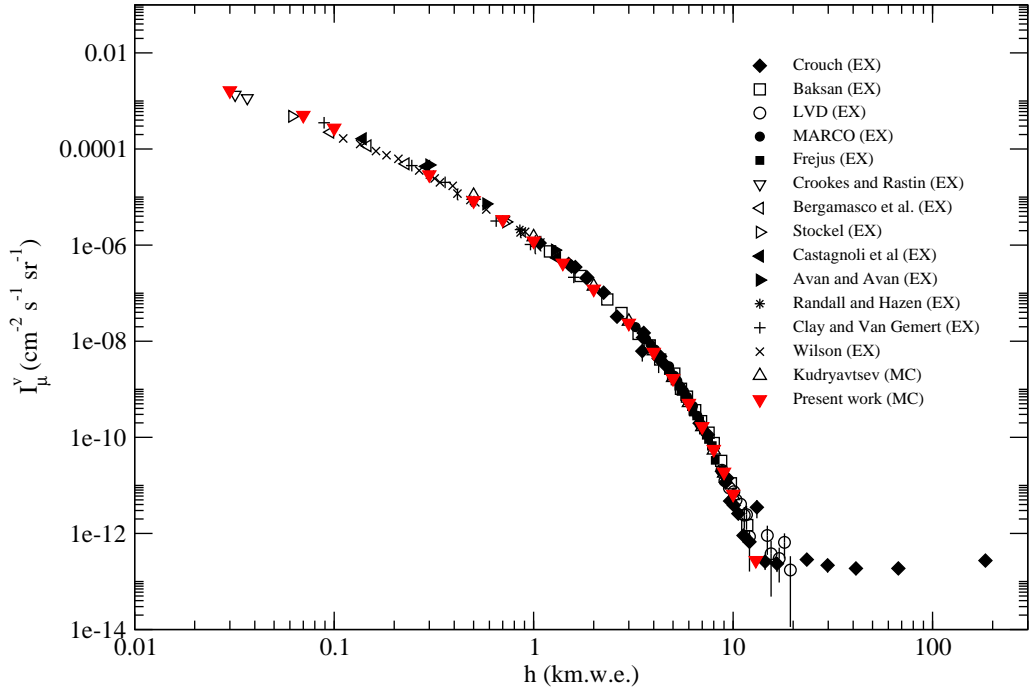}
\caption{\label{cr1}
Average vertical muon intensity $I^v_\mu(h)$ versus vertical
depth $h$ beneath a flat surface in standard rock.  Experimental and simulated
data are labeled by EX and MC respectively.  The experimental data of the flat
surface overburden are taken from
References~\cite{gaisser,bugaev} and the simulated data by Kudryavtsev
{\em et al.} from Reference`\cite{ku3}.  The number of simulated events per
data point in this figure is $N=10^6$.  The set of simulated data labeled
as ``Present work'' is generated from the modified Gaisser parameterization
of the surface muon intensity in Eqs.~(\ref{rc})--(\ref{a}).}
\end{figure}

\begin{figure}
\includegraphics[scale=0.55]{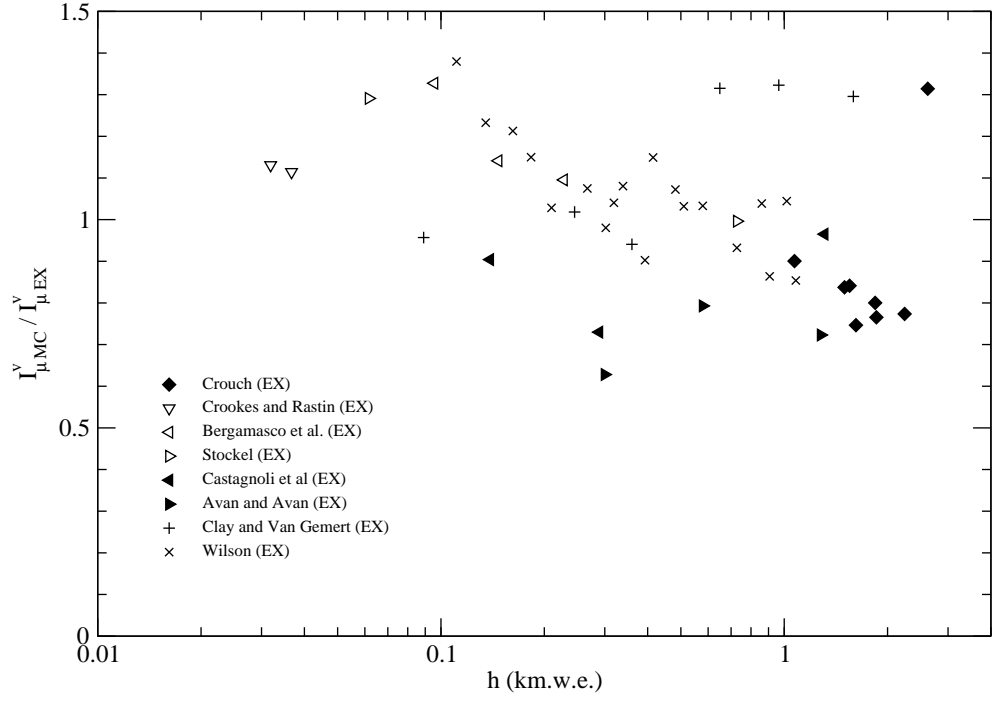}
\caption{\label{ratiov}
Ratio of simulated vertical muon intensity $I^v_\mu(h)_{MC}$
over the experimental vertical muon intensity $I^v_\mu(h)_{EX}$
versus shallow depth $h$ beneath a flat surface in standard rock.
The experimental data of the flat surface overburden are taken from
References~\cite{gaisser,bugaev}.  The number of simulated events per data
point in this figure is $N=10^6$.  The simulated data are generated from
the modified Gaisser parameterization incorporating
Eqs.~(\ref{rc})--(\ref{a}).}
\end{figure}

Vertical muon intensity and average energy are easily computed as
\ba
I^v_\mu(h)&\simeq&{\ln10\,(\log E_u-\log E_l)\over N}
\sum_{\{i\}}E_{\mu0i}\,
{dN_{\mu0}(E_{\mu0i},\cos\theta^\star_i=1)\over d E_{\mu0i}\, d\Omega},
\label{im2}\\
E^v_\mu(h)&\simeq&{\ln10\,(\log E_u-\log E_l)\over N\,I^v_\mu(h)}
\sum_{\{i\}}E_{\mu i}E_{\mu0i}\,
{dN_{\mu0}(E_{\mu0i},\cos\theta^\star_i=1)\over d E_{\mu0i}\, d\Omega}.
\label{im3}
\ea
Vertical muon intensity in standard rock simulated with Eq.~(\ref{im2}) is
compared against experimental data in Figs.~\ref{cr1} and \ref{ratiov}.  In
reality, standard rock does not exist and is generally not an exact match of
real rock profiles in real experiments.  When measurements are converted from
real rocks to standard rock, there are always some questions regarding the
accuracy of the conversion schemes.  For this reason, Figs.~\ref{cr1} and
\ref{ratiov} should not be taken as absolute tests of the accuracy of
\texttt{MUSIC} and the present integration
method but merely a relative point of reference.  Despite of the question of
the accuracy of the standard rock experimental data, it is shown in
Fig.~\ref{ratiov} that the ratios of calculated and experimental vertical
intensity scatter symmetrically around unity so that the simulated results
are said to agree with experiments at large.  It should be noted that vertical
intensity is merely an approximate test and is not the central focus of the
present work.  Integrated intensity on the other hand is really what is needed
for the calculation of muon overburden in realistic calculations.
Integrated muon intensity and average energy can be implemented in a similar
way as
Eqs.~(\ref{im2})--(\ref{im3}), 
\be
J_\mu\simeq{\Omega\ln10\,(\log E_u-\log E_l)\over N}
\sum_{\{i\}}E_{\mu0i}\,
{dN_{\mu0}(E_{\mu0i},\cos\theta^\star_i)\over d E_{\mu0i}\, d\Omega},
\label{jf1}
\ee
\be
E_\mu\simeq
{\Omega\ln10\,(\log E_u-\log E_l)\over N\,J_\mu(h)}
\sum_{\{i\}}E_{\mu i}\,E_{\mu0i}\,
{dN_{\mu0}(E_{\mu0i},\cos\theta^\star_i)\over d E_{\mu0i}\,d\Omega},
\label{jf2}
\ee
where $\Omega$ is the solid angle over the integrated hill profile.
The average muon energy $E_\mu$ can be organized into $M=500$
bins along $\log E_\mu$.  The subscript $j$ denotes the $j$-th bin.  The
numerical implementation of Eq.~(\ref{dist}) is
\be
{dJ_{\mu}\over dE_{\mu j}}\simeq
{\Omega M\over N}\sum^{N(E_{\mu j})}_{i=1}
{E_{\mu0 i}\over E_{\mu j}}
\,{dN_{\mu0}(E_{\mu0i},\cos\theta^\star_i)\over d E_{\mu0i}\,d\Omega}.
\label{adist}
\ee
Information of the survival probability is hidden in $N(E_{\mu j})$
that gives the number of surviving muons in the $j$-th bin.  As a
consistency check,
\be
J_\mu\simeq{\ln10(\log E_u-\log E_l)\over M}\sum^M_{j=1}E_{\mu j}\,
{dJ_\mu(h)\over dE_{\mu j}},
\label{jh4}
\ee
and
\be
E_\mu\simeq{\ln10(\log E_u-\log E_l)\over M\tilde J_\mu(h)}
\sum^M_{i=1}E_{\mu i}^2
{dJ_\mu(h)\over dE_{\mu j}}
\label{jh5}
\ee
must agree with those obtained by Eq.~(\ref{jf1}) and (\ref{jf2}).  Angular
and double differential distributions are constructed in similar ways.
For instance, the $\cos\theta$ distribution can be constructed as
\be
{dJ_\mu\over d\cos\theta_j}={\Omega\over2}\,{M\ln10(\log E_u-\log E_l)
\over N}\sum^{N(\cos\theta_j)}_{i=1}E_{\mu0i}
\,{dN_{\mu0}(E_{\mu0i},\cos\theta^\star_j)\over d E_{\mu0i}\,d\Omega}.
\label{dcz}
\ee
The factor $\Omega/2$ in Eq.~(\ref{dcz}) gives the proper normalization
so that the integration over $-1\le\cos\theta\le1$ gives the
correct solid angle $\Omega$.  Similarly the $\phi$ distribution is given as
\be
{dJ_\mu\over d\phi_j}={\Omega\over2\pi}\,{M\ln10(\log E_u-\log E_l)\over N}
\sum^{N(\phi_j)}_{i=1}E_{\mu0i}
\,{dN_{\mu0}(E_{\mu0i},\cos\theta^\star_i)\over d E_{\mu0i}\,d\Omega}.
\label{dph}
\ee
In the case of Eq.~(\ref{dph}), the normalization factor is $\Omega/2\pi$
so that the integration around $0\le\phi<2\pi$ gives the correct solid angle
$\Omega$.  There is a subtlety involving the $\theta$ distribution.  Since
$\cos\theta$ (not $\theta$) is uniformly generated in the present method,
uniform binning in $\theta$ leads to the wrong distribution.  The correct
bin width must be inversely proportional to $\cos\theta$ or, more precisely
speaking, equals $M/(N\cos\theta)$.  The factor of $1/\cos\theta$ exactly
cancels the factor
of $\cos\theta$ of the Jacobian so that the $\theta$ distribution is
\be
{dJ_\mu\over d\theta_j}={\Omega\over\pi}\,{M\ln10(\log E_u-\log E_l)\over N}
\sum^{N(\theta_j)}_{i=1}E_{\mu0i}
\,{dN_{\mu0}(E_{\mu0i},\cos\theta^\star_j)\over d E_{\mu0i}\,d\Omega}.
\label{dth}
\ee
Double differential distributions of various kinds can be constructed using
the same logic.

\begin{figure}
\includegraphics[scale=0.6]{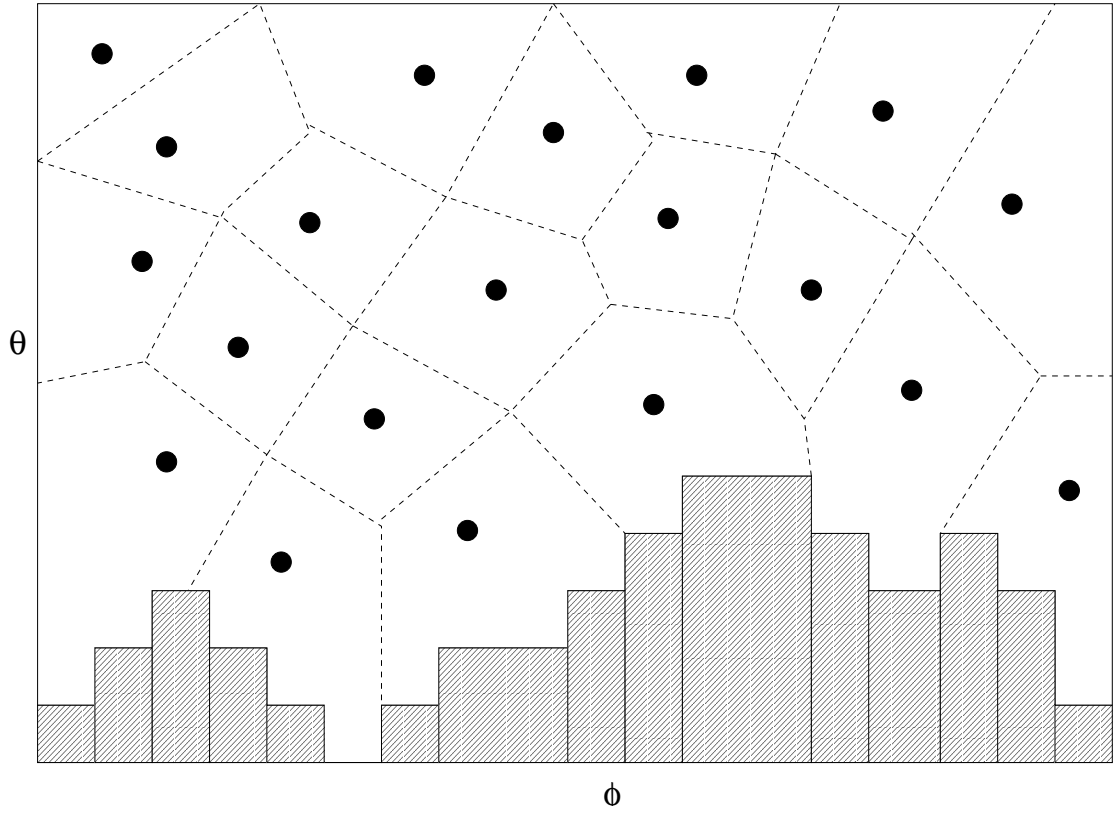}
\caption{\label{lattice}
An illustration of the binning strategy of slant depth $X$ in the $\theta-\phi$
space.  The bars represents regions of the solid angle corresponding to the
edges of a 3D topographical map and is blocked from the random generation of
$\theta$ and $\phi$.  The black dots represent the original lattice sites from
a latticized hill profile.  The dotted lines partition the remaining solid
angle into regions of nearest neighbors.}
\end{figure}

\begin{figure}
\includegraphics[scale=0.55]{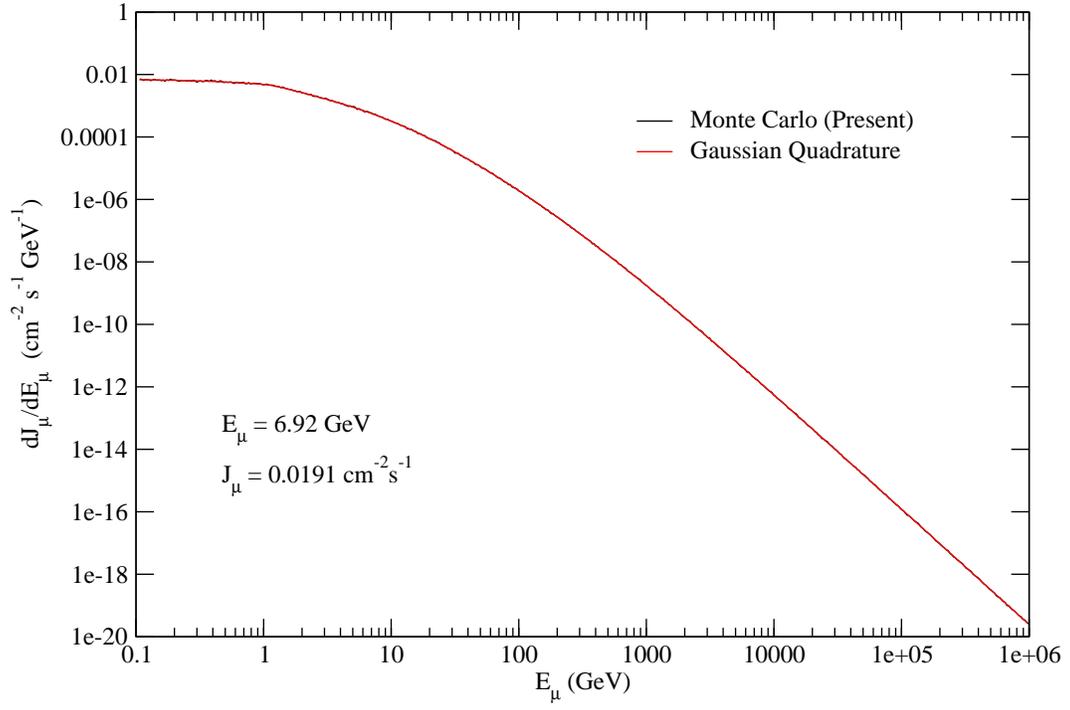}
\caption{\label{ge}
Integrated muon flux versus muon energy on ground level.  The angular
integration is taken over the the entire hemisphere.  Experimental data
support the feature that the energy spectrum for $E_\mu<1$~GeV is almost
flat~\cite{gaisser}.}
\end{figure}

\begin{figure}
\includegraphics[scale=0.55]{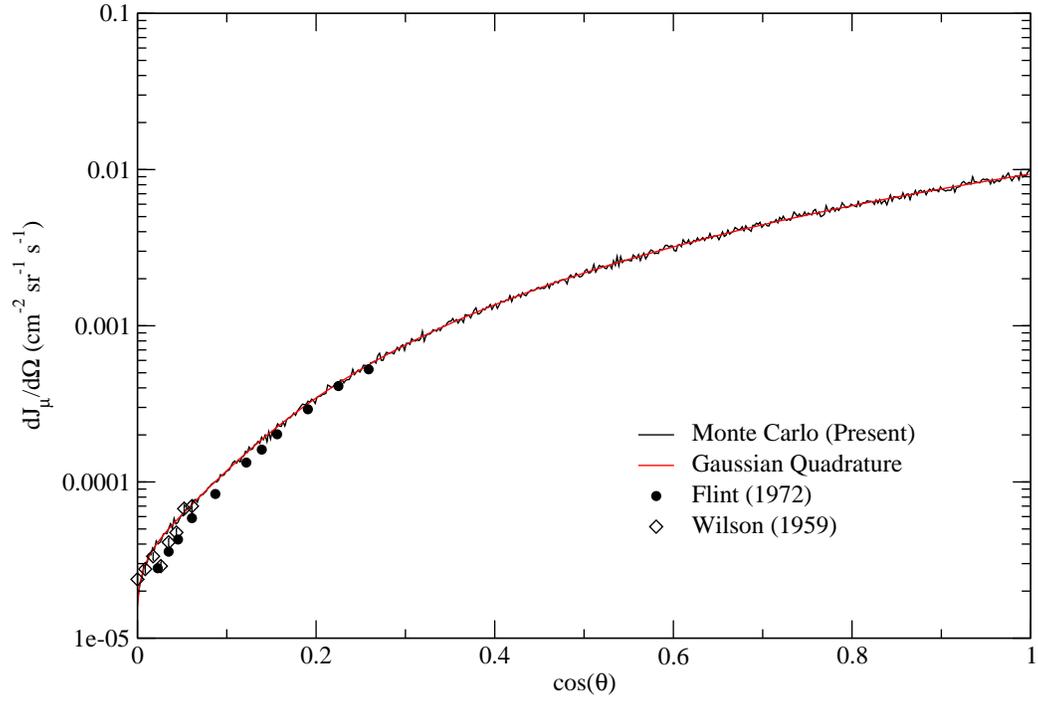}
\caption{\label{gc}
Integrated muon flux versus the zenith angle on ground level.  The energy
integration is taken over the range 0.106--$10^6$~GeV.  The experimental data
are taken from References~\cite{flint,wilson}.}
\end{figure}

Exact calculation of the slant depth $X$ is generally impossible for
any arbitrary latticized
hill profile.  Fortunately simulated energy loss by \texttt{MUSIC} is not very
sensitive to small changes in $X$ so that an approximation can be made.
Figure~\ref{lattice} illustrates the binning strategy of $X$ in the
$\theta-\phi$ space.  The idea is to partition the solid angle into regions of
nearest neighbors.  Each region has the same value of $X$.  Evenly generated
$\theta$ and $\phi$ pick out an approximate value of $X$ in the corresponding
nearest neighborhood.  There is a certain amount of computation overhead
in pre-processing the partitions.  When the number of simulated events is
sufficiently large, the overhead of partitioning is still more cost-effective
than a real-time search per event.
Due to the irregularity of the hill profile, any given
differential solid angle in the upper hemisphere may traverse disconnected
parts of the hill profile.  For this reason, the code must incorporate a
mechanism to pick out the appropriate slant depth $X$.  It can be easily
implemented by simply keeping only the minimum value of $X$ for any given grid
in a latticized $\theta-\phi$ space.  On ground level,
muons do not need to be propagated by \texttt{MUSIC} so that Eqs.~(\ref{is})
and (\ref{es}) can be integrated by Gaussian quadrature and by setting
$P(E_{\mu0},X,\theta^\star,\phi)=1$ in the integrand.
Figures~\ref{ge}--\ref{gc} and Table~\ref{gt} show that results generated by
the present Monte Carlo method agree exactly with those calculated by the
Gaussian quadrature method.

Assuming that pairs of $\cos\theta$, $\phi$ are uniformly generated $N$ times
and that the generation is truly random, the hemispherical $\Omega$ would have
been partitioned into $N$ segments uniformly.  In that case
$d{\mathbf A}\cdot\hat r=A_0\,\hat r\cdot\hat r_0/N$ so that the integral
of the differential projected surface areas is simply the sum of the segments
corresponding to the surviving muons.  In other words, for any given
$\hat r$, $\int d{\mathbf A}\cdot\hat r=
A_0\,\langle P\rangle\,\hat r\cdot\hat r_0$ where $\langle P\rangle$ is the
pseudo survival probability of muons.  A real survival probability of
muons can only be computed by generating muons according to the sea level muon
distribution and by propagating them through rock.  In the present method,
muons are generated uniformly in $E_{\mu0}$, $\cos\theta$ and $\phi$ so that
$\langle P\rangle$ does not have any natural meaning other than the ratio
of the surviving muons to the generated muons.  Na\"ively $\langle P\rangle$
may be set to the ratio of the number of surviving muons $n$ to the generated
muons $N$ according to this definition.  However a more careful look reveals
that $n$ increases as the muon energy threshold $E_l$ is raised in the
simulation.  The reason is simply that more highly energetic muons generally
have a better chance of surviving through rock.
For this reason, a proper definition of $\langle P\rangle$
must be independent of $E_l$ and is founded to be
\be
\langle P\rangle={n\over N}\,{\log E_u-\log E_l\over\log E_u-\log m_\mu}.
\label{p}
\ee
$\langle P\rangle$ defined in Eq.~(\ref{p}) is effectively independent of
$E_l$ because $n$ varies inversely with $\log E_l$.  On the other hand,
$n$ decreases when $E_u$ decreases because less highly energetic muons are
generally less likely to survive through rock.  Unfortunately
there is no simple way to rescale $\langle P\rangle$ in this case.  For an
accurate calculation of $R_\mu(h)$, it is recommended that $E_u$ is kept
at $10^6$~GeV.  With
$\langle P\rangle$ in place, Eq.~(\ref{micro}) can be implemented as
\be
R_\mu\simeq{A_0\,\langle P\rangle\,\Omega\ln10\,(\log E_u-\log E_l)\over N}
\sum_{\{i\}}\hat r_i\cdot\hat r_{0i}\,E_{\mu0i}\,
{dN_{\mu0}(E_{\mu0i},\cos\theta^\star_i)\over d E_{\mu0i}\, d\Omega}.
\label{ra}
\ee
The dot products $\hat r_i\cdot\hat r_{0i}$ are generated randomly inside
the entire detector volume and not just at the center.  This strategy renders
a fairer sampling of the detector geometry.  On the other hand, the hill
profile is defined with respect to the origin which is normally set at the
center of the detector because the generation of slant depth $X$ is not easily
managed when the origin moves.  Since \texttt{MUSIC} is relatively insensitive
to small change in $X$ and the size of the detector is generally small
compared to the mountain profile, generation of $X$ from the center of the
detector is adequate.

\begin{table}
\caption{Integrated muon intensity $J_\mu$ and energy $E_\mu$
versus vertical depth $h$ from a flat surface in standard rock.  Results
labeled ``Sheffield'' are taken from Reference~\cite{ku3} that
uses the original Gaisser parameterization and \texttt{MUSIC}.
Results labeled ``KSU'' present this work using the
modified Gaisser parameterization in Eqs.~(\ref{rc})--(\ref{a}) and
\texttt{MUSIC}.  The initial
muon energy for vertical depth $(300\le h\le2000)$~mwe is
$(0.106<E_{\mu0}\le10^6)$~GeV and for $(2000< h\le10000)$~mwe is
$(0.106<E_{\mu0}\le10^7)$~GeV.  The number of simulated events is $10^6$.}
\begin{ruledtabular}
\begin{tabular}{lrrrr}
& Sheffield & KSU & Sheffield & KSU \\
$h$~(mwe) & $J_\mu$ $\rm(cm^{-2}s^{-1})$ & $J_\mu$ $\rm(cm^{-2}s^{-1})$ 
& $E_\mu$ (GeV) & $E_\mu$ (GeV) \\
\tableline
500   &  $1.70\times10^{-5}$ & $1.71\times10^{-5}$ & 97  & 97 \\
1000  &  $2.20\times10^{-6}$ & $2.21\times10^{-6}$ & 157 & 158 \\
2000 & $1.81\times10^{-7}$ & $1.81\times10^{-7}$ & 236 & 236 \\
3000 & $2.94\times10^{-8}$ & $2.95\times10^{-8}$ & 285 & 284 \\
4000 & $6.33\times10^{-9}$ & $6.34\times10^{-9}$ & 316 & 313 \\
5000 & $1.58\times10^{-9}$ & $1.57\times10^{-9}$ & 337 & 339 \\
6000 & $4.30\times10^{-10}$ & $4.21\times10^{-10}$ & 351 & 345 \\
7000 & $1.24\times10^{-10}$ & $1.26\times10^{-10}$ & 361 & 365 \\
8000 & $3.73\times10^{-11}$ & $3.61\times10^{-11}$ & 369 & 356 \\
9000 & $1.15\times10^{-11}$ & $1.14\times10^{-11}$ & 375 & 373 \\
10000 & $3.65\times10^{-12}$ & $3.61\times10^{-12}$ & 380 & 363
\end{tabular}
\end{ruledtabular}
\label{cr}
\end{table}

\begin{table}
\caption{Chemical composition of the Ikenoyama and Ardennes rock in elemental
percentage.  The Ardennes rock composition is the average of several samples.
The CHOOZ rock data are approximate values only.  Details are documented in an
internal note~\cite{baldini}.}
\begin{ruledtabular}
\begin{tabular}{lrr}
& Ikenoyama & Ardennes \\
Chemical formula & \% & \%\\
\tableline
$\rm Si\,O_2$ & 60.70 & 58 \\
$\rm Ti\,O_2$ & 0.31 & \\
$\rm Al_2\,O_3$ & 17.39 & 19 \\
$\rm Fe_2\,O_3$ & 1.10 & \\
$\rm Fe\,O$ & 1.22 & 17 \\
$\rm Mn\,O$ & 0.15 & \\
$\rm Mg\,O$ & 0.93 & 4 \\
$\rm Ca\,O$ & 6.00 & \\
$\rm Na_2\,O$ & 6.42 & \\
$\rm K_2\,O$ & 3.47 & 2 \\
$\rm P_2\,O_5$ & 0.18 & \\
$\rm H_2\,O$ & 0.97 & \\
$\rm S$ & 0.01 & \\
$\rm C\,O_2$ & 0.96 &
\end{tabular}
\end{ruledtabular}
\label{chem}
\end{table}

\begin{table}
\caption{Average muon energy $E_\mu$, muon flux $J_\mu$, the muon rate
inside the tank $R_\mu^t$ and the muon rate inside the inner detector
volume $R_\mu^i$ for Super-K, KamLAND and CHOOZ.  The inner detector
of Super-K is a cylinder and that of KamLAND is a balloon.  The muon rate
inside the CHOOZ inner detector is not simulated.}
\begin{ruledtabular}
\begin{tabular}{lrrrr}
Site & $E_\mu$ (GeV) & $J_\mu$ $\rm(cm^{-2}s^{-1})$ & $R_\mu^t$ (Hz) &
$R_\mu^i$ (Hz) \\
\tableline
Super-K & $271\pm2$ & $(1.48\pm0.04)\times10^{-7}$
& $2.438\pm0.004$ & $1.828\pm0.003$ \\
KamLAND & $268\pm2$ & $(1.70\pm0.05)\times10^{-7}$
& $0.676\pm0.001$ & $0.246\pm0.001$ \\
CHOOZ & $60.6\pm0.4$ & $(6.12\pm0.07)\times10^{-5}$
& $30.5\pm0.2$ & -
\end{tabular}
\end{ruledtabular}
\label{tab}
\end{table}

\begin{figure}
\includegraphics[scale=0.6]{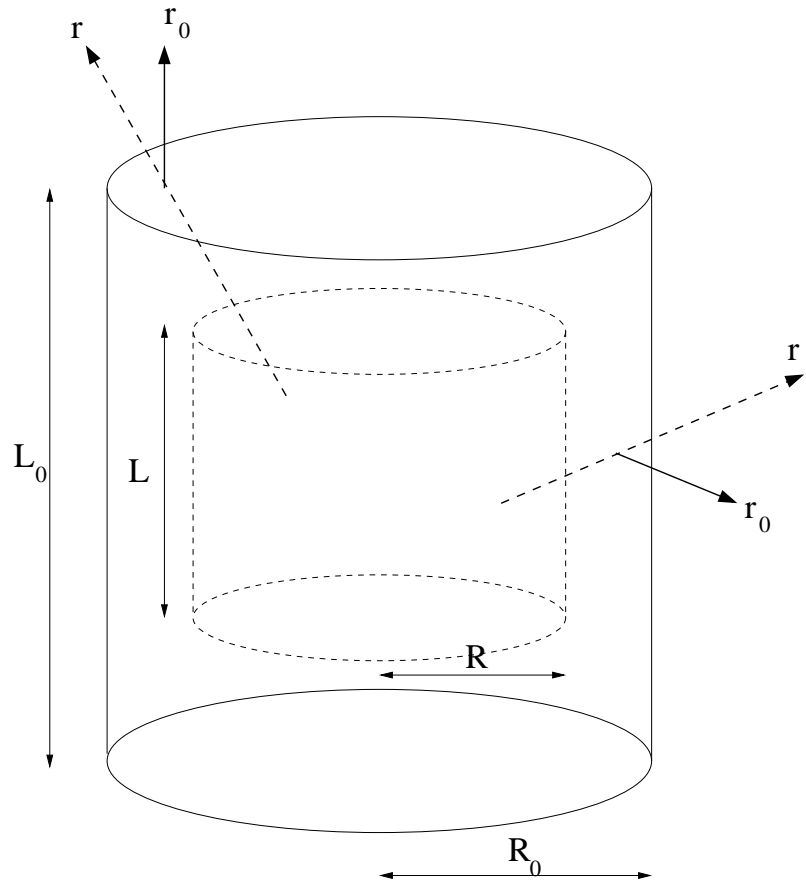}
\caption{\label{vol}
A sketch of a vertical cylindrical detector.  The inner volume is indicated
by dotted lines which is taken to be cylindrical for Super-K and CHOOZ but
spherical for KamLAND.}
\end{figure}

\begin{table}
\caption{Comparisons of muon flux and average energy on the ground level.  The
experimental
values of muon flux is taken from Reference~\cite{grieder}.  The momentum
cut-off of the muon flux measurement is 0.35~GeV.  The low total energy cut-off
of the present calculations is 0.106~GeV.  The quoted experimental
value~\cite{gaisser} of vertical muon energy is 4~GeV.  The simulated value is
4.19~GeV.}
\begin{ruledtabular}
\begin{tabular}{lll}
Method & $J_\mu$ $\rm(cm^{-2}\,s^{-1})$ & $E_\mu$ (GeV) \\
\tableline
Monte Carlo & $1.91\times10^{-2}$ & 6.92 \\
Gaussian Quadrature & $1.90\times10^{-2}$ & 6.95 \\
Experiments & $(1.90\pm0.12)\times10^{-2}$ & 
\end{tabular}
\end{ruledtabular}
\label{gt}
\end{table}

\end{document}